\documentclass[lettersize,journal]{IEEEtran}
\usepackage{amsmath,amsfonts}
\usepackage{algorithmic}
\usepackage{algorithm}
\usepackage{array}
\usepackage[caption=false,font=normalsize,labelfont=sf,textfont=sf]{subfig}
\usepackage{textcomp}
\usepackage{stfloats}
\usepackage{url}
\usepackage{verbatim}
\usepackage{graphicx}
\usepackage{cite}
\usepackage[hidelinks]{hyperref}
\usepackage{cleveref}

\crefname{figure}{figure}{figures}
\Crefname{figure}{Fig.}{Fig.}
\begin{document}

\title{Electromagnetic World Model for 6G: A Unified Framework for \\Joint Environment Reconstruction and  Channel Prediction}

\author{ Yizhu Zhao,~\IEEEmembership{Student Member,~IEEE,} Li Yu,~\IEEEmembership{Member,~IEEE,}  Jianhua Zhang,~\IEEEmembership{Fellow,~IEEE,} \\ Yuxiang Zhang,~\IEEEmembership{Member,~IEEE,} Zhen Zhang,~\IEEEmembership{Member,~IEEE,} Guangyi Liu,~\IEEEmembership{Member,~IEEE} 
        % <-this % stops a space
% <-this % stops a space
\thanks{This work was supported by the National Natural Science Foundation of China (62401084 and 62525101), the National Key R\&D Program of China (2023YFB2904801), the Mobile Information Networks-National Science and Technology Major Project (2025ZD1304600), and Beijing University of Posts and Telecommunications-China Mobile Communications Group Co.,Ltd. Joint Institute. (Corresponding author: Li Yu and Jianhua Zhang.)}
\thanks{Yizhu Zhao, Li Yu, Jianhua Zhang, and Yuxiang Zhang are with the State Key Laboratory of Networking and Switching Technology, Beijing University of Posts and Telecommunications, Beijing 100876, China (e-mail: \{zhaoyizhu, li.yu, jhzhang, zhangyx\}@bupt.edu.cn).}
\thanks{Zhen Zhang is with the Inner Mongolia Key Laboratory of Intelligent Communication and Sensing and Signal Processing, Inner Mongolia University, Hohhot 010021, China (e-mail: zhenzhang@imu.edu.cn).}
\thanks{Guangyi Liu is with China Mobile
Research Institute, Beijing 100053, China (e-mail: liuguangyi@chinamobile.com).}}



\maketitle

\begin{abstract}
The integration of sensing, communication, and intelligence is becoming a key enabler for sixth generation (6G) wireless systems, where intelligent terminals are expected to simultaneously support efficient link establishment and reliable environmental sensing. However, existing studies mainly exploit sensing information or communication information to address a single task, such as channel prediction or environment reconstruction. Motivated by the shared dependence of optical and radio-frequency signals on the surrounding environment, we propose the electromagnetic world model (EMWM), the first unified framework for joint environment reconstruction and channel prediction. EMWM learns a common electromagnetic representation with the potential to provide a modeling foundation for 6G tasks. Specifically, partial channel state information (CSI) and multi-view red-green-blue (RGB) images are encoded into CSI and visual tokens and jointly processed by a hierarchical world-model backbone with local and global aggregation. Based on the learned representation, a mixture-of-experts (MoE)-based CSI prediction head reconstructs the complete CSI, while a depth prediction head estimates multi-view depth maps that are further converted into three-dimensional (3D) point clouds. Moreover, a large-scale multi-modal dataset is constructed based on a campus digital twin. Experimental results show that EMWM outperforms conventional neural network and large language model (LLM) baselines in both CSI prediction and environment reconstruction, achieving a squared generalized cosine similarity (SGCS) of 0.9699 for CSI prediction while demonstrating robustness across different signal-to-noise ratio (SNR) conditions and zero-shot generalization at 28 GHz.
\end{abstract}


\section{Introduction}
\IEEEPARstart{T}{he} sixth generation (6G) is anticipated to usher in a new era where everything is sensed, connected, and intelligent \cite{ref1}. 
With the integration of advanced sensors and AI technologies, 6G is expected to empower intelligent terminals such as autonomous vehicles to achieve robust environmental sensing and efficient link establishment. 
In this context, future 6G systems are expected to increasingly exploit heterogeneous multi-modal information to support integrated sensing and communication (ISAC) and digital twin channel (DTC) \cite{ref2}.
However, the realization of this vision in 6G still entails substantial challenges from both the communication and sensing perspectives \cite{ref3}. 
On the one hand, although the increasing number of antennas in massive multi-input multi-output (MIMO) systems can significantly enhance spectrum efficiency \cite{ref4,ref5}, it also leads to prohibitively high pilot overhead for accurate channel state information (CSI) acquisition \cite{ref6}. 
On the other hand, for intelligent terminals such as vehicles, accurate depth information is indispensable for 3D sensing \cite{ref7}, whereas the widespread deployment of LiDAR remains constrained by its high cost compared with ordinary cameras \cite{ref8}. 
Therefore, various AI-enabled approaches have been proposed to address these communication and sensing challenges.

Recently, there have been a lot of outstanding works on channel prediction  \cite{ref9,ref10,ref11,ref12,ref13,ref14,ref15,ref16,ref17,ref18,ref19,ref20}, including CSI prediction and beam prediction, aiming to reduce the
overhead. For CSI prediction, \cite{ref9} proposes a CNN-based method by exploiting the spatial correlation between different antenna pairs, while \cite{ref10} comprehensively investigates AI-based channel extrapolation in the time, frequency, and space domains. In \cite{ref11}, a Doppler-adaptive channel estimation method based on a mixture-of-experts (MoE) architecture is proposed for complex mobile scenarios, and \cite{ref12} further exploits multi-view images to capture wireless environment information (WEI) for CSI prediction. Environment semantics has also been introduced for beam prediction, where \cite{ref13} develops a beam prediction method based on propagation environment semantics for non-line-of-sight (NLOS) scenarios and \cite{ref14} proposes an environment semantics aided framework for mmWave beam prediction.

In particular, large language models (LLMs) have also been incorporated into channel prediction due to their strong generalization capability. LLM4CP~\cite{ref15} is proposed to improve channel prediction accuracy and generalization, while \cite{ref16} develops a prompt-enabled large AI model for CSI feedback by incorporating environmental knowledge. ChannelGPT~\cite{ref17} further integrates multi-modal information, particularly multi-view images, for multi-task channel prediction. For beam prediction, \cite{ref18} reformulates the task as time-series
forecasting, while \cite{ref19} proposes LLM-MM, a multi-modal beam prediction framework based on mixture-of-experts with low-rank adaptation (LoRA) to improve robustness across diverse scenarios. ChannelDS~\cite{ref20} further integrates multi-view images and position information to enable few-shot beam prediction.

In addition, recent studies have also explored the integration of wireless communication and AI for environment reconstruction \cite{ref21,ref22,ref23,ref24,ref25,ref26,ref27,ref28}, including 3D point cloud reconstruction and depth prediction. In \cite{ref21}, a model-driven deep learning framework jointly performs radio map learning and 3D environment reconstruction, while \cite{ref22} proposes a deep-learning-based multinode ISAC 4D environmental reconstruction method with uplink-downlink cooperation. A noise-sparsity-aware diffusion model is further introduced in \cite{ref23} to denoise and densify coarse point clouds. Similarly, Wi-Fi CSI can be exploited for 3D point cloud reconstruction. CSI2PC~\cite{ref24} generates point clouds from measured CSI, and \cite{ref25} leverages temporal CSI amplitude and phase information for 3D point cloud reconstruction. By integrating camera and radar data, AI-enabled depth prediction has also been investigated. For example, \cite{ref26} fuses camera images with mmWave radar point clouds, TacoDepth~\cite{ref27} introduces graph-based radar structure extraction and pyramid-based fusion, and \cite{ref28} combines ISAC signals with camera images through multi-user selection and multi-modal fusion.

In fact, existing studies have yet to establish a unified AI-enabled framework for communication and sensing. Most current efforts either focus on exploiting sensing information, such as WEI \cite{ref29}, for channel prediction, or on leveraging communication information, such as CSI, for environment reconstruction. 
This separation overlooks the intrinsic coupling between the physical environment and wireless propagation, since environmental factors such as blockage and clutter density can significantly affect wireless propagation characteristics \cite{ref41}. 
Furthermore, although LLMs have recently been introduced into channel prediction, their pre-training paradigm is still primarily established on large-scale natural language corpora rather than visual sensing data containing WEI. As a result, they do not natively provide an adequate mechanism for capturing propagation characteristics, including scene geometry, blockage conditions, and scatterer distribution. This limitation makes it difficult for LLM-based approaches to effectively model the relationship between environmental information and channel propagation, thereby restricting their suitability for wireless channel prediction.

\begin{figure*}[t]
    \centering
    \includegraphics[width=0.85\textwidth]{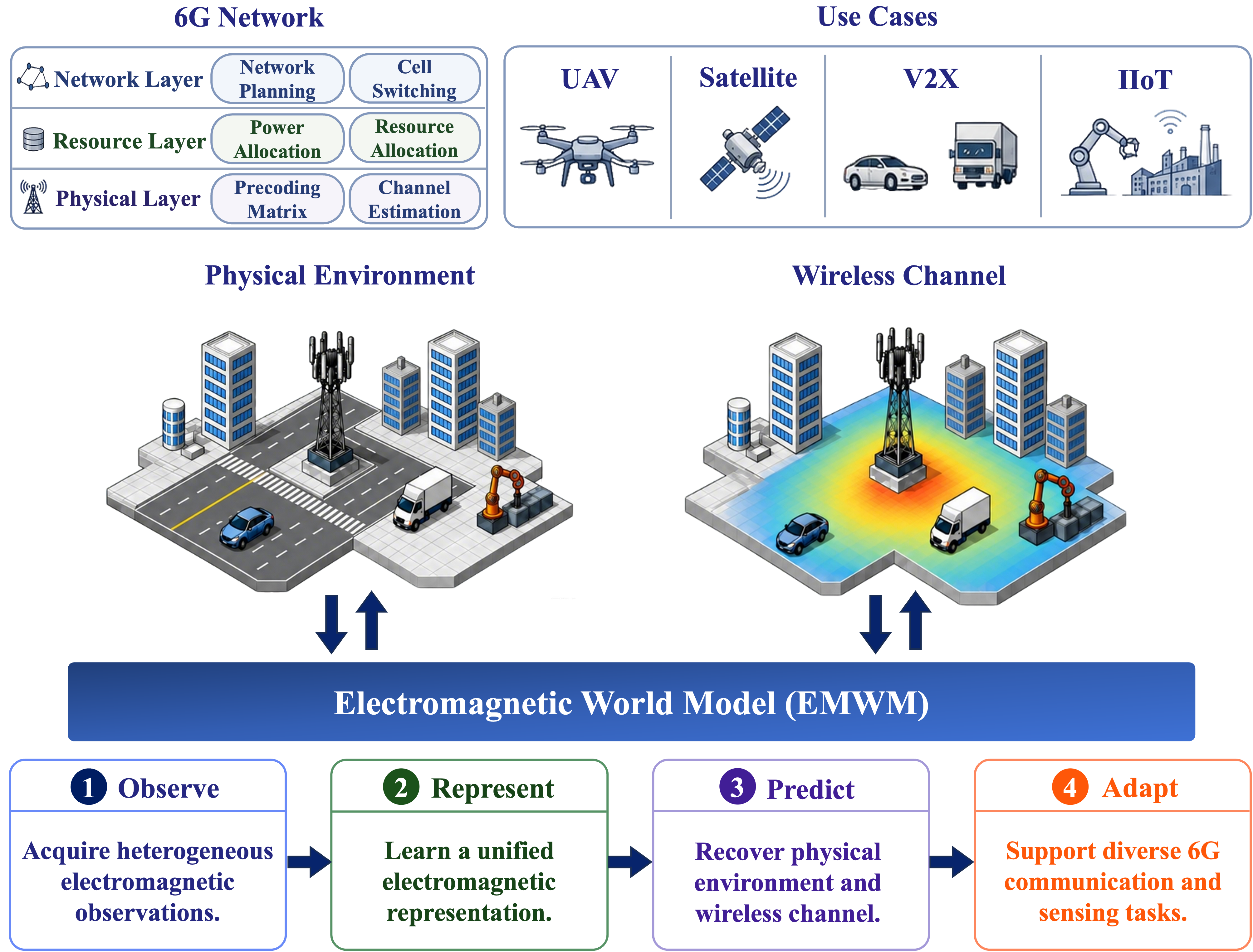}
    \caption{Illustration of the electromagnetic world model for 6G.}
    \label{fig:emwm_illustration}
\end{figure*}

Compared with LLMs, world models offer a more suitable paradigm for this challenge, since they learn compact representations from sensory inputs and predict the underlying attributes or dynamics of the physical world \cite{ref30}.  In particular, Voyager \cite{ref31} is proposed as a world-consistent video diffusion framework that jointly generates aligned red-green-blue (RGB) and depth sequences, supports direct 3D scene reconstruction, and is naturally capable of video depth prediction. Recent studies have begun to explore world models in wireless communications, for example, by using world foundation models to predict future video frames \cite{ref40}. However, existing world models are mainly developed upon optical observations, such as images, and therefore remain insufficient for wireless communication systems. Although optical observations provide rich information about scene geometry, object boundaries, and semantic structures, existing world models do not natively support radio-frequency observations as model inputs. From a broader physical perspective, optical observations and radio-frequency observations can both be regarded as electromagnetic observations from different frequency bands, since they are carried by electromagnetic waves and implicitly reflect physical-world attributes. Therefore, how to establish a unified framework that can fuse heterogeneous electromagnetic observations for joint environment reconstruction and channel prediction warrants further investigation.

Motivated by this, we propose the electromagnetic world model (EMWM) for 6G, a unified framework for joint environment reconstruction and channel prediction. As illustrated in \Cref{fig:emwm_illustration}, EMWM follows a four-stage paradigm. Specifically, it acquires heterogeneous electromagnetic observations, learns a unified electromagnetic representation, predicts complete states of the physical environment and wireless channel, and supports diverse 6G tasks. By capturing the intrinsic coupling between the physical environment and wireless channel, EMWM provides a common modeling foundation across the network, resource, and physical layers and can be applied to diverse 6G application scenarios.

In this paper, we instantiate and validate the proposed EMWM through joint 3D environment reconstruction and  CSI prediction. Specifically, the proposed EMWM takes multi-view images and partial CSI as inputs, transforms them into visual and CSI tokens, respectively, and feeds these tokens into the world-model backbone to predict complete CSI and per-view depth maps. By further leveraging camera parameters, the predicted depth maps can be converted into 3D point clouds. In this way, the proposed EMWM reduces pilot overhead while enabling explicit 3D environment reconstruction. The main contributions of this paper are summarized as follows.
\begin{itemize}
    \renewcommand\labelitemi{\raisebox{-0.32em}{\Large\textbullet}}
    \item We make the first attempt to define the electromagnetic world model for 6G as a unified framework for joint environment reconstruction and channel prediction. Different from existing world models mainly developed upon optical observations, the proposed EMWM acquires heterogeneous electromagnetic observations, learns a unified electromagnetic representation, predicts complete states of the physical environment and wireless channel, and supports diverse 6G tasks.
    
    \item We develop a concrete EMWM architecture for joint 3D environment reconstruction and channel prediction. Specifically, uniformly sampled partial CSI is converted into CSI tokens and incorporated into the backbone together with visual tokens, so that optical and radio-frequency electromagnetic observations can be jointly modeled. Moreover, a MoE-based CSI prediction head is designed for complete CSI prediction, while LoRA is employed to efficiently fine-tune the  backbone architecture of the world model.

    \item A large-scale multi-modal dataset is constructed based on the digital twin of Beijing University of Posts and Telecommunications (BUPT). Generated through autonomous driving simulation software and ray-tracing software, the dataset contains multi-view images, corresponding depth information, and complete CSI labels. It contains 36 routes and 36{,}197 samples in total, thereby providing a comprehensive data foundation for training and evaluating the proposed  unified framework.

    \item Comprehensive experimental results demonstrate that the proposed EMWM outperforms both LLM-based methods and conventional small models for joint environment reconstruction and channel prediction. Specifically, for environment reconstruction, EMWM achieves lower depth prediction error and produces more accurate depth maps for subsequent 3D point cloud  reconstruction. For channel prediction, EMWM achieves lower NMSE and higher SGCS, while also showing robustness under different signal-to-noise ratio (SNR) conditions and cross-frequency zero-shot generalization.
\end{itemize}

The rest of this paper is organized as follows. Section~\ref{sec:system}  introduces the system model and formulates the considered problem. Section~\ref{sec:framework} presents the proposed EMWM framework. Section~\ref{sec:simulation} provides the simulation settings and performance analysis. Finally, Section~\ref{sec:conclusion} concludes this paper.

\textit{Notations}: Boldface lowercase letters $\mathbf{x}$ and boldface uppercase letters $\mathbf{X}$ denote vectors and matrices, respectively. The $m$-th row and the $n$-th column of $\mathbf{X}$ are denoted by $\mathbf{X}_{m,:}$ and $\mathbf{X}_{:,n}$, respectively, and $\mathbf{X}_{m,n}$ denotes the element at the $m$-th row and the $n$-th column. The operators $(\cdot)^*$, $(\cdot)^T$, and $(\cdot)^H$ denote conjugate, transpose, and conjugate transpose, respectively. The sets $\mathbb{R}^N$ and $\mathbb{C}^N$ denote $N$-dimensional real and complex vector spaces, respectively. $\mathcal{S}$ denotes an index set, and $\mathcal{P}$ denotes a point cloud set. The notation $[\cdot;\cdot]$ denotes concatenation.

\section{System Model and Problem Formulation}\label{sec:system}
\subsection{System Model}
\begin{figure}[!t]
\centerline{\includegraphics[width=0.95\linewidth]{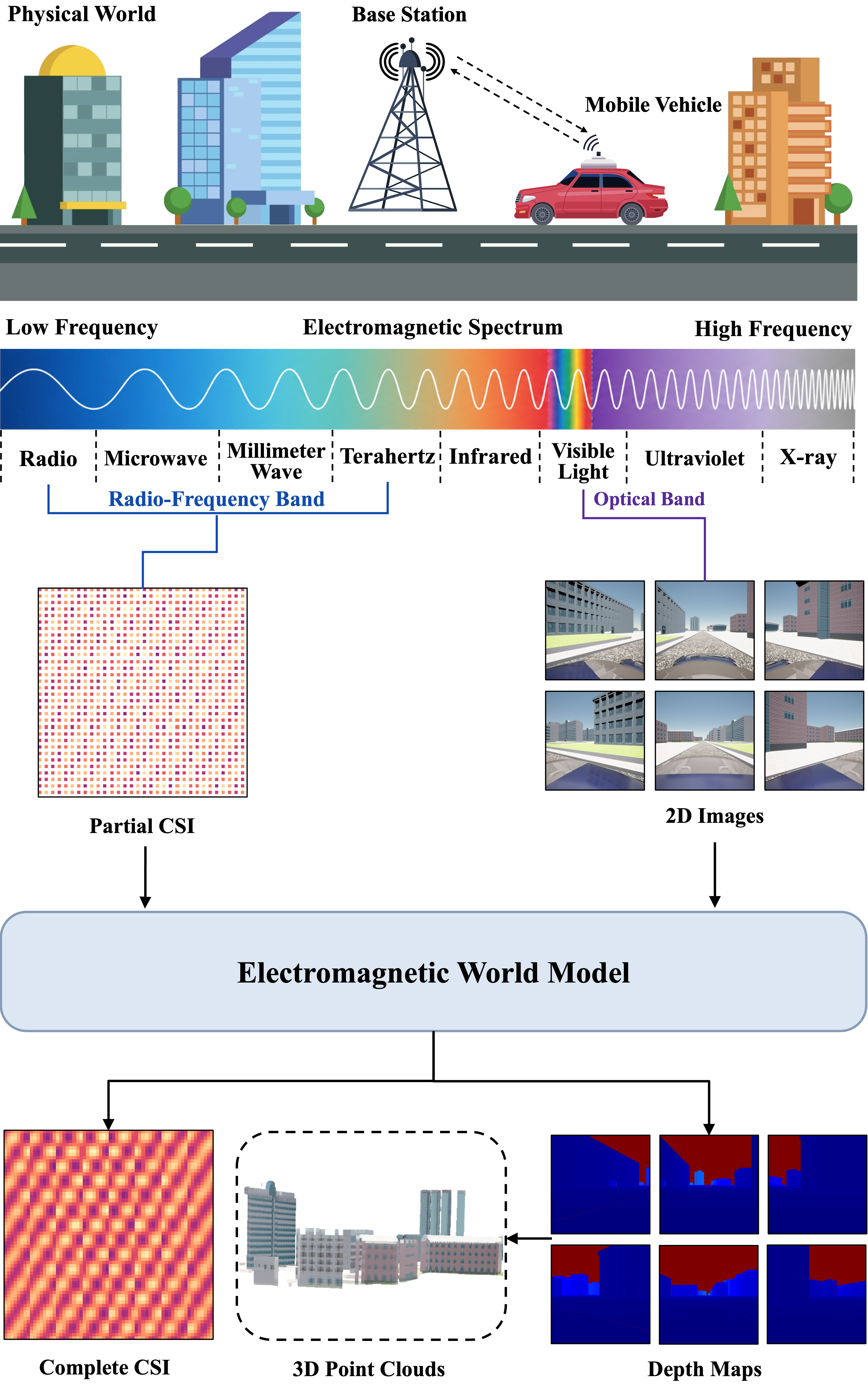}}
\caption{System model for environment reconstruction and channel prediction.}
\label{fig:fig1}
\end{figure}

Consider an outdoor urban single-cell frequency division duplexing (FDD) multiple-input single-output (MISO) system, consisting of a base station (BS) and a mobile station (MS). 
As shown in \Cref{fig:fig1}, the BS is equipped with a uniform planar array (UPA) of $\textit{N}_{b}$ antennas to communicate with a single-antenna mobile vehicle. In addition, the mobile vehicle is equipped with multiple RGB cameras to capture multi-view images $\{\mathbf{I}_i\}_{i=1}^{N_c}$, where $\mathbf{I}_i \in \mathbb{R}^{W \times H \times 3}$, and $N_c$ denotes the number of camera viewpoints.

The communication system employs orthogonal frequency division multiplexing (OFDM) with $N_s$ subcarriers for data transmission, and the downlink channel at the $k$-th subcarrier can be expressed as
\begin{equation}
\mathbf{h}[k]=\sum_{l=1}^{L} \alpha_{l} e^{-j2\pi f_{k}\tau _{l} + j\mathit{\psi } _{l}} \mathbf{a}\left ( \theta _{l} , \phi_{l}  \right ),
\end{equation}
where $\alpha_l$, $\tau_l$, and $\psi_l$ denote the attenuation, delay, and phase shift of the $l$-th propagation path, respectively, and $f_k$ is the frequency of the $k$-th subcarrier. Additionally, $L$ denotes the total number of multipath components. The parameters $\theta_l$ and $\phi_l$ denote the azimuth and elevation angle of departure corresponding to the $l$-th path, respectively, and $\mathbf{a}(\theta_l,\phi_l)$ represents the transmit array response vector \cite{ref32}. Assuming that the antenna spacing is equal to half a wavelength, $\mathbf{a}(\theta_l,\phi_l)$ can be given by
\begin{equation}
\begin{aligned}
\mathbf{a}(\theta_l, \phi_l) = &  \frac{1}{\sqrt{N_{b}^{h}N_{b}^{v}} } \left[ 1, \dots, e^{j\pi\left[ h \cos \left( \phi_l \right) + v \sin \left( \theta_l \right) \sin \left( \phi_l \right) \right]}, \right. \\
& \left. \dots, e^{j\pi\left[ \left(N_{b}^{h}-1\right) \cos \left( \phi_l \right) +  \left(N_{b}^{v}-1\right) \sin \left( \theta_l \right) \sin \left( \phi_l \right) \right]} \right]^{T} ,
\end{aligned}
\end{equation}
where $N_b = N_b^{h} N_b^{v}$, in which $N_b^{h}$ and $N_b^{v}$ denote the numbers of antenna elements arranged along the horizontal and vertical directions, respectively.

By stacking the channel vectors over all subcarriers, the whole  channel response matrix in the frequency domain can be written as
\begin{equation}
\mathbf{H} = \left[ \mathbf{h}[1], \mathbf{h}[2], \ldots, \mathbf{h}[N_s] \right] \in \mathbb{C}^{N_b \times N_s}.
\end{equation}

To acquire channel information, the BS transmits downlink pilot symbols for channel estimation. Let $\mathbf{x}[k] \in \mathbb{C}^{N_b \times 1}$ denote the pilot signal transmitted by the BS on the $k$-th subcarrier. Then, the received pilot signal at the mobile vehicle can be expressed as
\begin{equation}
y[k] = \mathbf{h}^{H}[k]\mathbf{x}[k] + n[k],
\end{equation}
where $y[k] \in \mathbb{C}$ is the received signal at the mobile vehicle, and $n[k] \sim \mathcal{CN}(0,\sigma^2)$ denotes the additive white Gaussian noise (AWGN).
To reduce the pilot overhead, the BS employs a uniform sampling strategy over the spatial and frequency domains, such that downlink pilots are transmitted only on selected antennas and subcarriers.
 Specifically, let $\mathcal{S}_b \subseteq \{1,2,\ldots,N_b\}$ and $\mathcal{S}_s \subseteq \{1,2,\ldots,N_s\}$ denote the uniformly sampled antenna index set and  uniformly sampled subcarrier index set, respectively, where $|\mathcal{S}_b| = N_b^{p}$ and $|\mathcal{S}_s| = N_s^{p}$. Based on the noisy downlink pilot signals received on these sampled antennas and subcarriers, the mobile vehicle performs channel estimation, yielding the partial CSI matrix
\begin{equation}
\widetilde{\mathbf{H}}=\left[\tilde{\mathbf{h}}[1],\tilde{\mathbf{h}}[2],\ldots,\tilde{\mathbf{h}}[N_s^{p}]\right]\in\mathbb{C}^{N_b^{p}\times N_s^{p}},
\end{equation}
where $\tilde{\mathbf{h}}[q] \in \mathbb{C}^{N_b^p \times 1}$ denotes the estimated channel vector on the sampled antennas for the $q$-th sampled subcarrier, and $\widetilde{\mathbf{H}}$ represents the estimated partial CSI corresponding to the sampled antennas and sampled subcarriers.

For subsequent processing, $\widetilde{\mathbf{H}}$ is mapped into a matrix of the same dimension as the full CSI through zero padding, resulting in a matrix $\mathbf{H}^{p} \in \mathbb{C}^{N_b \times N_s}$, whose entries can be expressed as
\begin{equation}
\mathbf{H}^{p}_{m,n} =
\begin{cases}
\widetilde{\mathbf{H}}_{u,q}, & m = \mathcal{S}_b(u),\; n = \mathcal{S}_s(q),\\
0, & \text{otherwise},
\end{cases}
\end{equation}
where $u \in \{1,2,\ldots,N_b^{p}\}$ and $q \in \{1,2,\ldots,N_s^{p}\}$ denote the indices in the partial CSI matrix. Thus, $\mathbf{H}^{p}$ preserves the estimated CSI at the sampled antennas and subcarriers, while assigning zeros to all unobserved entries.

\subsection{Problem Formulation}
In this paper, our primary objective is to predict physical-world attributes by leveraging heterogeneous electromagnetic observations acquired from the mobile vehicle. 
Specifically, the partial CSI and multi-view images are regarded as radio-frequency and optical observations, respectively, and are jointly exploited to predict the complete CSI and multi-view depth maps. Furthermore, the predicted multi-view depth maps are  converted into a 3D point cloud representation of the surrounding environment using the camera  parameters.

To this end, we aim to design a unified framework for joint environment reconstruction and channel prediction. The framework is deployed on the vehicle side to perform inference. The objective of this framework is to learn a mapping function $\Psi_{\omega}$ that establishes the relationship between the input electromagnetic observations and the desired physical-world attributes. Specifically, the input data consist of the  partial CSI matrix $\widetilde{\mathbf{H}} \in\mathbb{C}^{N_b^{p}\times N_s^{p}}$ and the multi-view images $\{\mathbf{I}_i\}_{i=1}^{N_c}$, while the outputs consist of the predicted complete CSI matrix $\widehat{\mathbf{H}} \in \mathbb{C}^{N_b \times N_s}$ and the predicted multi-view depth maps $\{\widehat{\mathbf{D}}_i\}_{i=1}^{N_c}$ with $\widehat{\mathbf{D}}_i \in \mathbb{R}^{W \times H}$. Accordingly, the mapping function can be expressed as
\begin{equation}
\Psi_{\omega}:\left(\widetilde{\mathbf{H}},\{\mathbf{I}_i\}_{i=1}^{N_c}\right)\rightarrow\left(\widehat{\mathbf{H}},\{\widehat{\mathbf{D}}_i\}_{i=1}^{N_c}\right),
\end{equation}
where $\omega$ denotes the trainable parameters of the proposed framework.

Furthermore, we introduce a deterministic geometric transformation based on the camera  parameters to convert the predicted multi-view depth maps into 3D point clouds, thereby enabling 3D environment reconstruction. Specifically, for the $i$-th camera viewpoint, the corresponding 3D point cloud can be expressed as
\begin{equation}
\widehat{\mathcal{P}}_{i}=\mathcal{T}\left(\widehat{\mathbf{D}}_{i},\mathbf{K}_{i}\right),
\end{equation}
where $\widehat{\mathcal{P}}_{i}$ denotes the 3D point cloud generated from the predicted depth map $\widehat{\mathbf{D}}_{i}$, $\mathbf{K}_{i}$ denotes the  matrix of the $i$-th camera, and $\mathcal{T}(\cdot)$ represents a fixed geometric transformation from the image plane to the 3D space. The final 3D point cloud representation is then obtained by aggregating the 3D points from all camera viewpoints.

\begin{figure*}[!t]
\centerline{\includegraphics[width=0.9\linewidth]{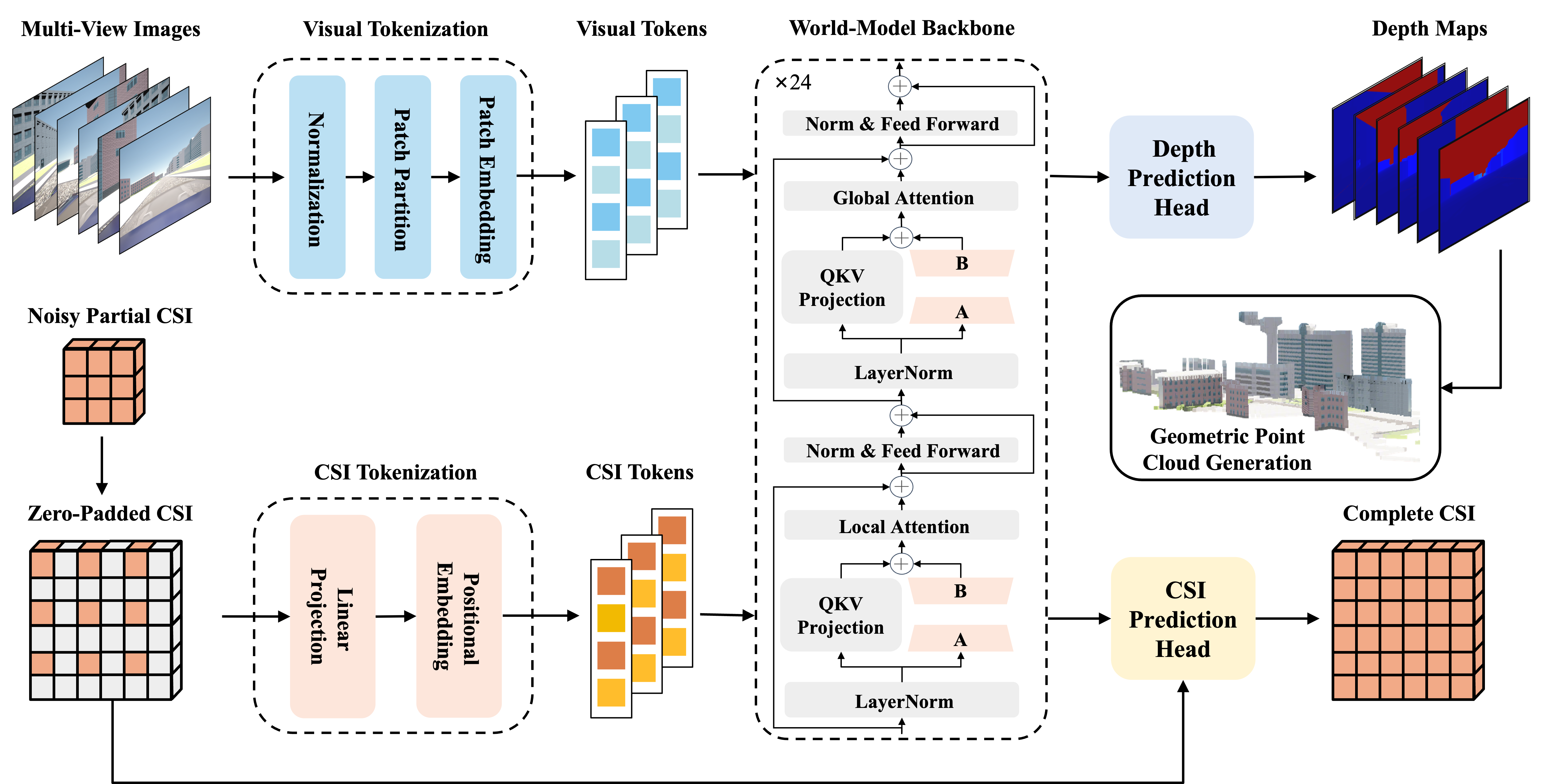}}
\caption{ The network design of the proposed EMWM.}
\label{fig:fig2}
\end{figure*}

\section{The Proposed EMWM Architecture}\label{sec:framework}
In this section, we present a concrete EMWM architecture for  joint environment reconstruction and channel prediction. To support multi-view geometric representation learning, the proposed framework leverages the HunyuanWorld-Mirror \cite{ref33} model released by Tencent Hunyuan. HunyuanWorld-Mirror provides unified geometric modeling capability to extract and aggregate multi-view visual features into geometrically consistent representations. Motivated by this property, we adopt its  backbone  and further extend the input token space with CSI tokens. In this way, the extended backbone can jointly capture scene geometry and wireless propagation characteristics, while supporting environment reconstruction and channel prediction through task-specific heads. 
As shown in \Cref{fig:fig2}, the framework consists of six modules: visual tokenization, CSI tokenization, world-model backbone, depth prediction head, MoE-based CSI prediction head, and geometric point cloud generation. 
The details of the network architecture and the optimization objectives are illustrated below.

\subsection{Network Architecture}
\subsubsection{Visual Tokenization Module}
For the multi-view RGB images $\{\mathbf{I}_i\}_{i=1}^{N_c}$, we employ the visual tokenization module to transform raw image observations into a token sequence. Specifically, mean-standard deviation normalization is first applied to each input image, which improves the convergence rate during model training. The normalization process can be mathematically represented as
\begin{equation}
\bar{\mathbf{I}}_i=\frac{\mathbf{I}_i-\boldsymbol{\mu}}{\boldsymbol{\sigma}},
\end{equation}
where $\boldsymbol{\mu}$ and $\boldsymbol{\sigma}$ denote the mean and standard deviation, respectively. The normalized image $\bar{\mathbf{I}}_i$ is then divided into $N_p$ regular patches, which are further mapped into a $d_m$ dimensional latent space through the patch embedding operator. Accordingly, the visual tokens of the $i$-th view $\mathbf{T}^{\mathrm{img}}_i \in \mathbb{R}^{N_p \times d_m}$ are formulated as
\begin{equation}
\mathbf{T}^{\mathrm{img}}_i=
\left[
\phi\!\left(\bar{\mathbf{I}}_i^{(1)}\right),
\phi\!\left(\bar{\mathbf{I}}_i^{(2)}\right),
\ldots,
\phi\!\left(\bar{\mathbf{I}}_i^{(N_p)}\right)
\right]^T,
\end{equation}
where $\bar{\mathbf{I}}_i^{(p)}$ denotes the $p$-th patch of the normalized image $\bar{\mathbf{I}}_i$, $\phi(\cdot)$ denotes the patch embedding operator. By collecting all viewpoints, the obtained visual tokens serve as the visual input to the world-model backbone. In this way, the raw multi-view images are converted into token representations that preserve local spatial structure and support subsequent feature aggregation.

\subsubsection{CSI Tokenization Module}
To integrate wireless propagation characteristics into the proposed framework, we further introduce the CSI tokenization module. Specifically, the obtained partial CSI is first expanded to the full dimension through zero padding, yielding
\begin{equation}
\mathbf{H}^{p}=\mathcal{Z}\!\left(\widetilde{\mathbf{H}}\right),
\end{equation}
where $\mathcal{Z}(\cdot)$ denotes the zero-padding operator that preserves the estimated entries in $\widetilde{\mathbf{H}}$ and fills the unobserved positions with zeros. 

Then, $\mathbf{H}^{p}$ is reformulated into a real-valued tensor by separating its real and imaginary parts, so that the complex CSI can be represented in a form suitable for neural processing. Based on this representation, a learnable linear projection is employed to map each complex CSI entry into a $d_m$ dimensional latent space. Furthermore, positional embeddings are introduced to preserve the structural arrangement of CSI over the antenna and subcarrier dimensions. Accordingly, the CSI tokens $\mathbf{T}^{\mathrm{csi}} \in \mathbb{R}^{N_bN_s\times d_m}$ are formulated as
\begin{equation}
\mathbf{T}^{\mathrm{csi}}
=
\left[
\psi\!\left(\mathbf{H}^{p}_{1,1}\right),
\psi\!\left(\mathbf{H}^{p}_{1,2}\right),
\ldots,
\psi\!\left(\mathbf{H}^{p}_{N_b,N_s}\right)
\right]^T
+
\mathbf{\Gamma}^{\mathrm{pos}},
\end{equation}
where $\psi(\cdot)$ denotes the linear CSI embedding operator and $\mathbf{\Gamma}^{\mathrm{pos}}$ denotes the positional embedding. In this way, the sparse CSI observations are transformed into CSI tokens and aligned with the visual tokens in the same latent space, thereby enabling the subsequent backbone to jointly aggregate visual and CSI features.

\subsubsection{World-Model Backbone Module}
The world-model backbone is responsible for jointly modeling visual tokens and CSI tokens within a unified latent space. Specifically, the visual tokens and CSI tokens are first concatenated to form a unified token sequence $\mathbf{T}\in\mathbb{R}^{(N_cN_p+N_bN_s)\times d_m}$, yielding
\begin{equation}
\mathbf{T}=
\left[
\mathbf{T}^{\mathrm{img}};
\mathbf{T}^{\mathrm{csi}}
\right].
\end{equation}

The backbone adopts a hierarchical transformer architecture consisting of 24 stacked stages, where each stage includes a local aggregation block followed by a global aggregation block. The feature transformation in each block is implemented via multi-head self-attention followed by a feed-forward network.

In the multi-head attention operator, for the $i$-th head, the query, key, and value matrices are first computed through linear projections. Rotary positional embedding (RoPE) is further applied to the query and key representations to encode spatial structure. The attention operation can be formulated as
\begin{equation}
\operatorname{ATTENTION}(\mathbf{Q}_i,\mathbf{K}_i,\mathbf{V}_i)=\mathbf{A}_i\mathbf{V}_i,
\end{equation}
\begin{equation}
\mathbf{A}_i=
\operatorname{Softmax}
\left(
\frac{(\mathbf{Q}_i^{\mathrm{r}})(\mathbf{K}_i^{\mathrm{r}})^{T}}{\sqrt{d_h}}
\right),
\end{equation}
where $\mathbf{Q}_i^{\mathrm{r}}$ and $\mathbf{K}_i^{\mathrm{r}}$ denote the query and key after RoPE, and $d_h=d_m/N_h$ is the dimension of each attention head with $N_h$ being the number of heads.

To support downstream prediction tasks, the backbone further outputs multi-level feature representations from several selected transformer layers. Specifically, for the $u$-th selected layer, the local feature representation and the global feature representation are concatenated along the feature dimension to form the output tokens $\mathbf{T}^{\mathrm{out}}_{u} \in
\mathbb{R}^{(N_cN_p+N_bN_s)\times 2d_m}$, which can be expressed as
\begin{equation}
\mathbf{T}^{\mathrm{out}}_{u}
=
\mathcal{F}_u(\mathbf{T})
=
\left[
\mathbf{T}^{\mathrm{loc}}_{u};
\mathbf{T}^{\mathrm{glo}}_{u}
\right],
\end{equation}
where $\mathbf{T}^{\mathrm{loc}}_{u}$ and $\mathbf{T}^{\mathrm{glo}}_{u}$ denote the local and global representations obtained from the transformation $\mathcal{F}_u(\cdot)$, respectively. By collecting features from multiple layers, the backbone produces a set of multi-level representations, which are subsequently utilized by the task-specific prediction heads.

To reduce computational cost during training, we adopt LoRA for efficient fine-tuning of the backbone. Assuming the pre-trained weight matrix $\mathbf{W}_0\in\mathbb{R}^{d_m\times d_m}$, the adapted weight is expressed as
\begin{equation}
\mathbf{W}=\mathbf{W}_0+\frac{\alpha}{r}\mathbf{B}\mathbf{A},
\end{equation}
where $\mathbf{A}\in\mathbb{R}^{r\times d_m}$ and $\mathbf{B}\in\mathbb{R}^{d_m\times r}$ are trainable low-rank matrices, while $r$ and $\alpha$ denote the rank and scaling factor, respectively. This design significantly reduces the number of trainable parameters while preserving the expressive capability of the backbone.

\subsubsection{Depth Prediction Head Module}
We adopt the original depth prediction head as implemented in HunyuanWorld-Mirror to perform dense depth prediction from the multi-level backbone representations. Specifically, for each selected layer \(u\), only the visual  tokens are extracted from \(\mathbf{T}^{\mathrm{out}}_{u}\). For each viewpoint, the corresponding visual tokens are processed independently. Accordingly, the input feature of the \(u\)-th level for the \(i\)-th viewpoint can be written as
\begin{equation}
\mathbf{F}_{u,i}
=
\mathcal{R}\!\left(
\mathcal{O}_{\mathrm{img}}\!\left(\mathbf{T}^{\mathrm{out}}_{u}\right)
\right)
\in
\mathbb{R}^{2d_m\times \frac{H}{P}\times \frac{W}{P}},
\end{equation}
where \(\mathcal{O}_{\mathrm{img}}(\cdot)\) extracts the visual tokens corresponding to the \(i\)-th viewpoint, \(\mathcal{R}(\cdot)\) denotes the reshape operator that maps the token sequence into a spatial feature map, and \(P=\sqrt{\frac{HW}{N_p}}\) denotes the patch size.

Based on these multi-level inputs, the depth prediction head first applies layer normalization and \(1\times1\) convolutional projection to each level to obtain task-specific feature representations. Then, scale alignment is performed through a set of resolution adjustment layers, including upsampling, identity mapping, and downsampling operations, so that multi-level features can be fused in a unified spatial resolution. Subsequently, a top-down refinement pathway composed of multiple fusion blocks is employed to progressively aggregate hierarchical information. Each fusion block integrates residual convolution units and interpolation operations, enabling effective multi-scale feature aggregation.

Finally, the fused feature map is processed by a lightweight convolutional head to produce depth predictions. Denoting the predicted depth map of the \(i\)-th viewpoint by \(\widehat{\mathbf{D}}_i\in
\mathbb{R}^{W\times H}\), the overall mapping can be expressed as
\begin{equation}
\widehat{\mathbf{D}}_i
=
\mathcal{H}_{\mathrm{dep}}\!\left(
\{\mathbf{F}_{u,i}\}_{u\in\mathcal{U}}
\right),
\end{equation}
where \(\mathcal{U}\) denotes the set of selected backbone stages and \(\mathcal{H}_{\mathrm{dep}}(\cdot)\) denotes the depth prediction function. In this way, the inherited depth prediction head converts multi-level token representations into multi-view dense depth maps for subsequent 3D environment reconstruction.

\begin{figure}[!t]
\centerline{\includegraphics[width=0.92\linewidth]{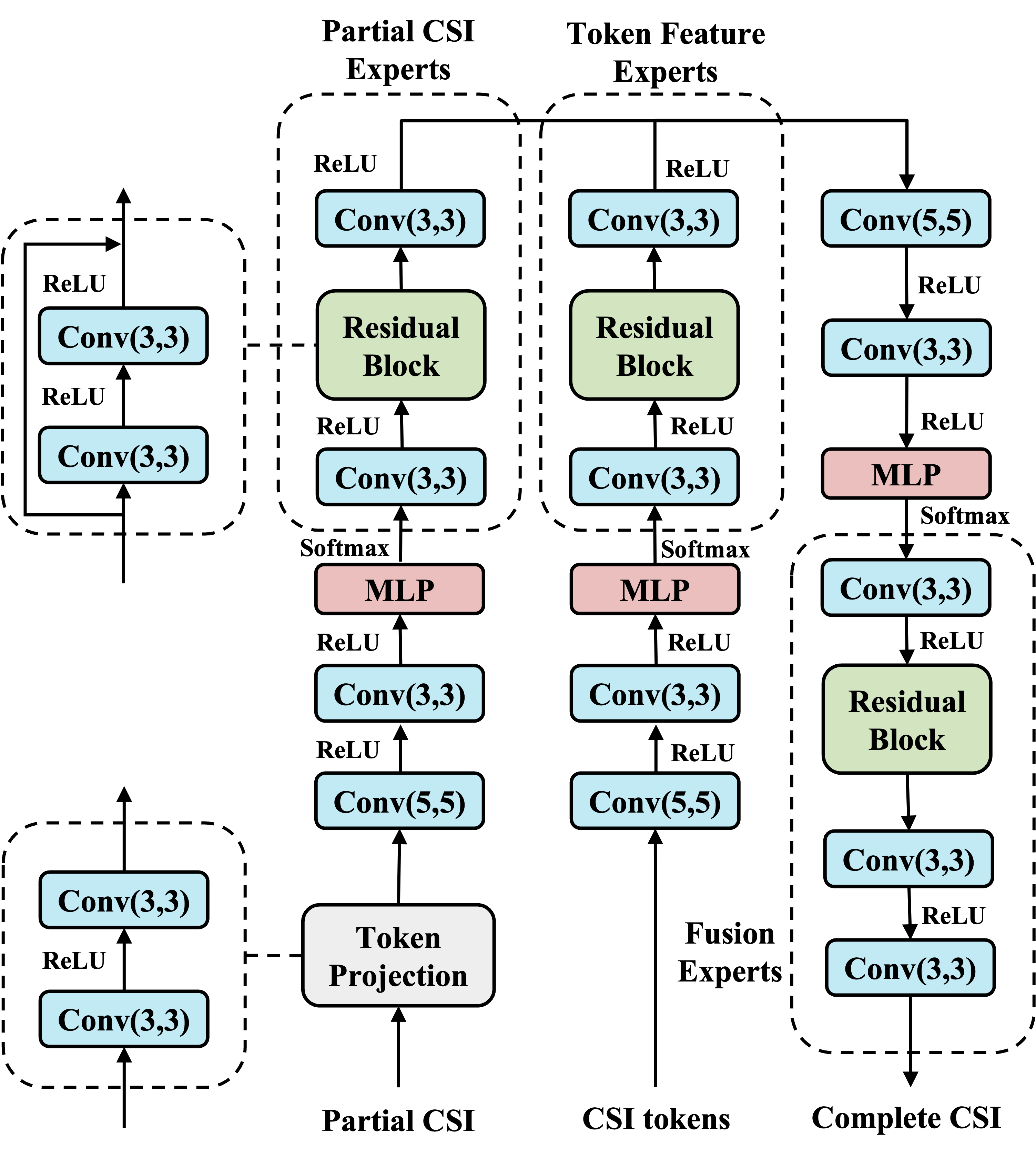}}
\caption{The architecture of the proposed MoE-based CSI prediction head.}
\label{fig:fig3}
\end{figure}

\subsubsection{MoE-Based CSI Prediction Head Module}
As illustrated in \Cref{fig:fig3}, we further introduce an MoE-based CSI prediction head in EMWM, which reconstructs the complete CSI by jointly leveraging the observed partial CSI and the CSI token representations extracted from the final backbone output.

Specifically, the input to the proposed head consists of two parts. The first part is the partial CSI matrix \(\mathbf{H}^p\), which is represented in the implementation by separating its real and imaginary parts into a real-valued tensor. The second part is constructed by extracting the CSI token part from the last backbone output. Let \(U\) denote the index of the last backbone stage. Then, the CSI token input $\mathbf{Z}_{U} \in
\mathbb{R}^{N_bN_s\times 2d_m}$ can be expressed as
\begin{equation}
\mathbf{Z}_{U}
=
\mathcal{O}_{\mathrm{csi}}\!\left(
\mathbf{T}^{\mathrm{out}}_{U}
\right)
,
\end{equation}
where \(\mathcal{O}_{\mathrm{csi}}(\cdot)\) extracts the CSI token component from the backbone output.

Based on the two inputs, the proposed CSI prediction head adopts a three-branch architecture, including a partial CSI branch, a token feature branch, and a fusion branch. Specifically, the CSI token representation \(\mathbf{Z}_{U}\) is first reshaped into a two-dimensional feature map and then projected by two convolutional layers into a compact token feature map, which provides structural guidance for CSI reconstruction. In parallel, the partial CSI branch takes the partial CSI matrix \(\mathbf{H}^p\), represented by its real and imaginary parts, as input, while the token feature branch takes the resulting token feature map as input. 

Both branches adopt gated mixture-of-experts for feature extraction. In each branch, a gate network first encodes the input feature map through convolutional layers and global average pooling, and then generates expert selection weights through a multilayer perceptron. Meanwhile, each expert consists of convolutional layers and residual blocks to extract spatial features. For an input feature map \(\mathbf{X}\), the corresponding MoE mapping can be written as
\begin{equation}
\mathcal{M}(\mathbf{X})
=
\sum_{k=1}^{K}
\pi_k(\mathbf{X})\,\mathcal{E}_k(\mathbf{X}),
\end{equation}
where \(K\) denotes the number of experts, \(\mathcal{E}_k(\cdot)\) denotes the \(k\)-th expert, and \(\pi_k(\mathbf{X})\) denotes the corresponding gate weight. The gate weight is computed as
\begin{equation}
\pi_k(\mathbf{X})
=
\frac{\exp\!\left(g_k(\mathbf{X})\right)}
{\sum_{j=1}^{K}\exp\!\left(g_j(\mathbf{X})\right)},
\end{equation}
where \(g_k(\mathbf{X})\) is the gating score generated by the gate network.

After that, the features extracted from the partial CSI branch and the token feature branch are concatenated along the feature dimension and fed into the fusion branch, which employs another MoE module to adaptively combine the two types of features and reconstruct the complete CSI. Denoting the predicted complete CSI by \(\widehat{\mathbf{H}}\), the output of the proposed head can be expressed as
\begin{equation}
\widehat{\mathbf{H}}
=
\mathcal{M}_{\mathrm{fus}}
\!\left(
\left[
\mathcal{M}_{\mathrm{par}}(\mathbf{H}^{p});
\mathcal{M}_{\mathrm{tok}}(\mathbf{Z})
\right]
\right),
\end{equation}
where \(\mathcal{M}_{\mathrm{par}}(\cdot)\), \(\mathcal{M}_{\mathrm{tok}}(\cdot)\), and \(\mathcal{M}_{\mathrm{fus}}(\cdot)\) denote the MoE mappings of the partial CSI branch, token feature branch, and fusion branch, respectively. In this way, the proposed head enables adaptive CSI reconstruction by jointly exploiting the partial CSI input and the CSI token representations extracted from the last backbone output.

\subsubsection{Geometric Point Cloud Generation Module}
\begin{figure}[htbp]
\centerline{\includegraphics[width=0.88\linewidth]{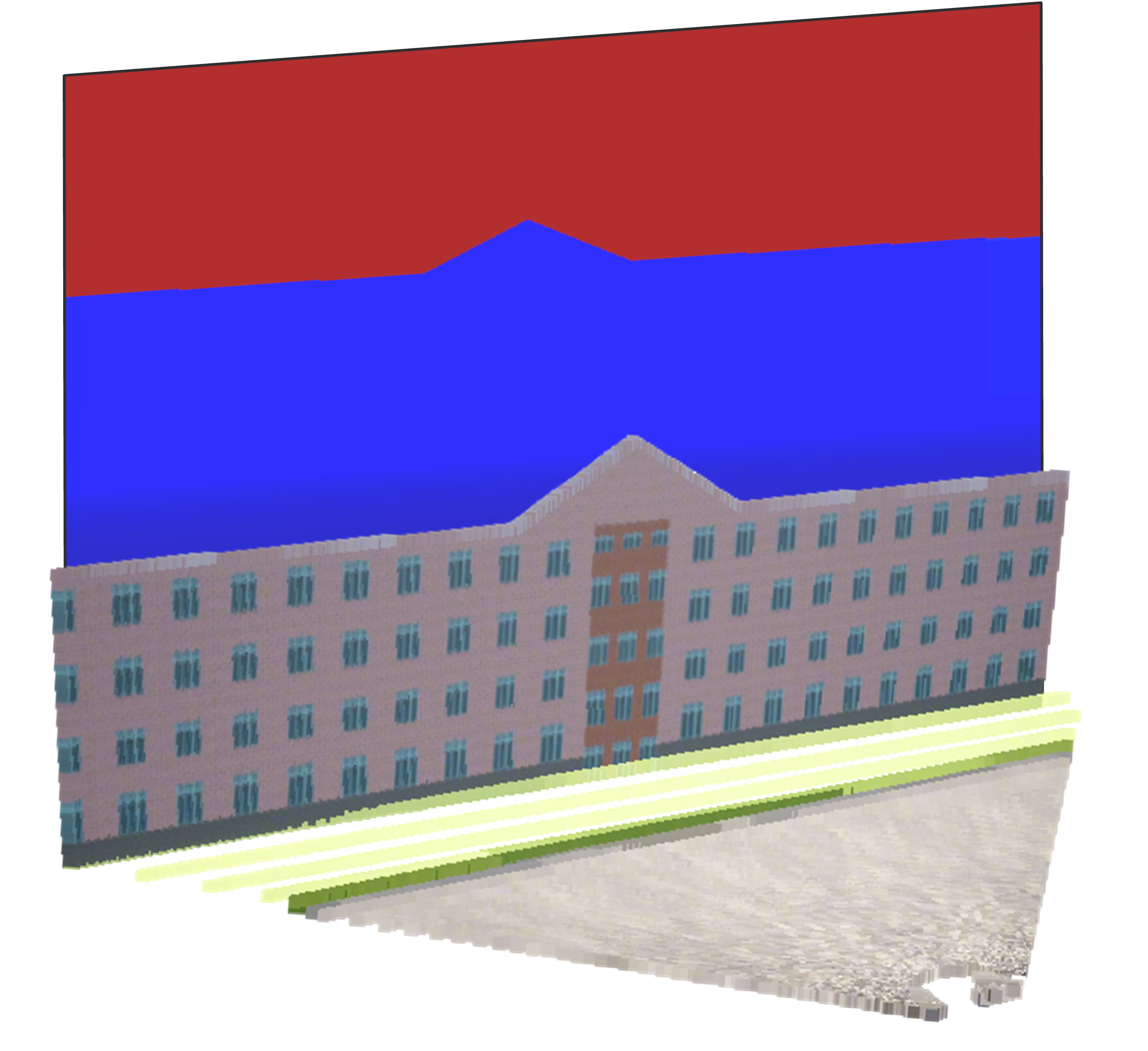}}
\caption{Geometric point cloud generation from depth maps.}
\label{fig:fig4}
\end{figure}

After obtaining the predicted multi-view depth maps \(\{\widehat{\mathbf{D}}_i\}_{i=1}^{N_c}\), we further perform geometric point cloud generation to reconstruct the surrounding 3D environment. As illustrated in \Cref{fig:fig4}, each depth map is first converted into a set of 3D points via camera  parameters. The final 3D point cloud is obtained by aggregating the point clouds from all viewpoints.
Specifically, for the \(i\)-th viewpoint, let \((u,v)\) denote the pixel coordinates and \(\widehat{\mathbf{D}}_i(u,v)\) denote the predicted depth value at that pixel. Based on the  matrix \(\mathbf{K}_i\), the corresponding 3D point in the camera coordinate system is obtained by back-projection as
\begin{equation}
\mathbf{p}_{i}(u,v)=
\begin{bmatrix}
\frac{(u-c_x)\widehat{\mathbf{D}}_i(u,v)}{f_x}\\[3pt]
\frac{(v-c_y)\widehat{\mathbf{D}}_i(u,v)}{f_y}\\[3pt]
\widehat{\mathbf{D}}_i(u,v)
\end{bmatrix},
\end{equation}

where \(f_x\) and \(f_y\) denote the focal lengths along the horizontal and vertical directions, respectively, and \(c_x\) and \(c_y\) denote the coordinates of the principal point, all determined by the  matrix \(\mathbf{K}_i\). By traversing all pixels in \(\widehat{\mathbf{D}}_i\), the 3D point cloud of the \(i\)-th viewpoint is generated as
\begin{equation}
\widehat{\mathcal{P}}_{i}
=
\left\{
\mathbf{p}_{i}(u,v)\,\middle|\,1\leq u\leq W,\;1\leq v\leq H
\right\}
.
\end{equation}

After that, the final reconstructed point cloud is obtained by aggregating the point sets generated from all viewpoints, and can be written as
\begin{equation}
\widehat{\mathcal{P}}
=
\bigcup_{i=1}^{N_c}\widehat{\mathcal{P}}_{i}.
\end{equation}

Algorithm~\ref{alg:depth2pc} summarizes the geometric procedure for converting the predicted multi-view depth maps into the reconstructed point cloud.

\begin{algorithm}[H]
\caption{Geometric Point Cloud Generation from Multi-View Depth Maps}
\label{alg:depth2pc}
\begin{algorithmic}
\STATE
\STATE {\textsc{GeneratePointCloud}}$(\{\widehat{\mathbf{D}}_i\}_{i=1}^{N_c}, \{\mathbf{K}_i\}_{i=1}^{N_c})$
\STATE \hspace{0.5cm} Initialize $\widehat{\mathcal{P}} \gets \emptyset$
\STATE \hspace{0.5cm} \textbf{for} $i = 1,2,\ldots,N_c$
\STATE \hspace{1.0cm} Initialize $\widehat{\mathcal{P}}_i \gets \emptyset$
\STATE \hspace{1.0cm} \textbf{for each pixel} $(u,v)$ \textbf{in} $\widehat{\mathbf{D}}_i$
\STATE \hspace{1.5cm} Compute $\mathbf{p}_{i}(u,v)$ from $\widehat{\mathbf{D}}_i(u,v)$ and $\mathbf{K}_i$
\STATE \hspace{1.5cm} $\widehat{\mathcal{P}}_i \gets \widehat{\mathcal{P}}_i \cup \{\mathbf{p}_{i}(u,v)\}$
\STATE \hspace{1.0cm} \textbf{end for}
\STATE \hspace{1.0cm} $\widehat{\mathcal{P}} \gets \widehat{\mathcal{P}} \cup \widehat{\mathcal{P}}_i$
\STATE \hspace{0.5cm} \textbf{end for}
\STATE \hspace{0.5cm} \textbf{return} $\widehat{\mathcal{P}}$
\end{algorithmic}
\end{algorithm}

\subsection{Optimization Objectives}
During the training process, the proposed EMWM is optimized in an end-to-end manner by jointly supervising CSI prediction and depth prediction. Let \(\mathcal{L}_{\mathrm{csi}}\) and \(\mathcal{L}_{\mathrm{dep}}\) denote the CSI prediction loss and the depth prediction loss, respectively. Then, the overall training objective can be expressed as
\begin{equation}
\mathcal{L}
=
\lambda_{\mathrm{csi}}\mathcal{L}_{\mathrm{csi}}
+
\lambda_{\mathrm{dep}}\mathcal{L}_{\mathrm{dep}},
\end{equation}
where \(\lambda_{\mathrm{csi}}\) and \(\lambda_{\mathrm{dep}}\) denote the weighting coefficients used to balance the contributions of the CSI prediction loss and the depth prediction loss.

For the CSI prediction task, to improve reconstruction quality in both the amplitude and phase domains while directly constraining the global complex error, we adopt a joint CSI loss function composed of amplitude MSE, phase MSE, and complex-domain normalized mean squared error (NMSE). The corresponding loss function is written as
\begin{equation}
\mathcal{L}_{\mathrm{csi}}
=
\mathcal{L}_{\mathrm{amp}}
+
\mathcal{L}_{\mathrm{pha}}
+
\mathcal{L}_{\mathrm{nmse}},
\end{equation}
where \(\mathcal{L}_{\mathrm{amp}}\), \(\mathcal{L}_{\mathrm{pha}}\), and \(\mathcal{L}_{\mathrm{nmse}}\) denote the amplitude reconstruction loss, phase reconstruction loss, and normalized mean squared error, respectively. Such a design enables the model to preserve both  phase information and the overall complex CSI structure.

\begin{figure}[!t]
\centering

\subfloat[]{%
    \includegraphics[height=0.195\textheight]{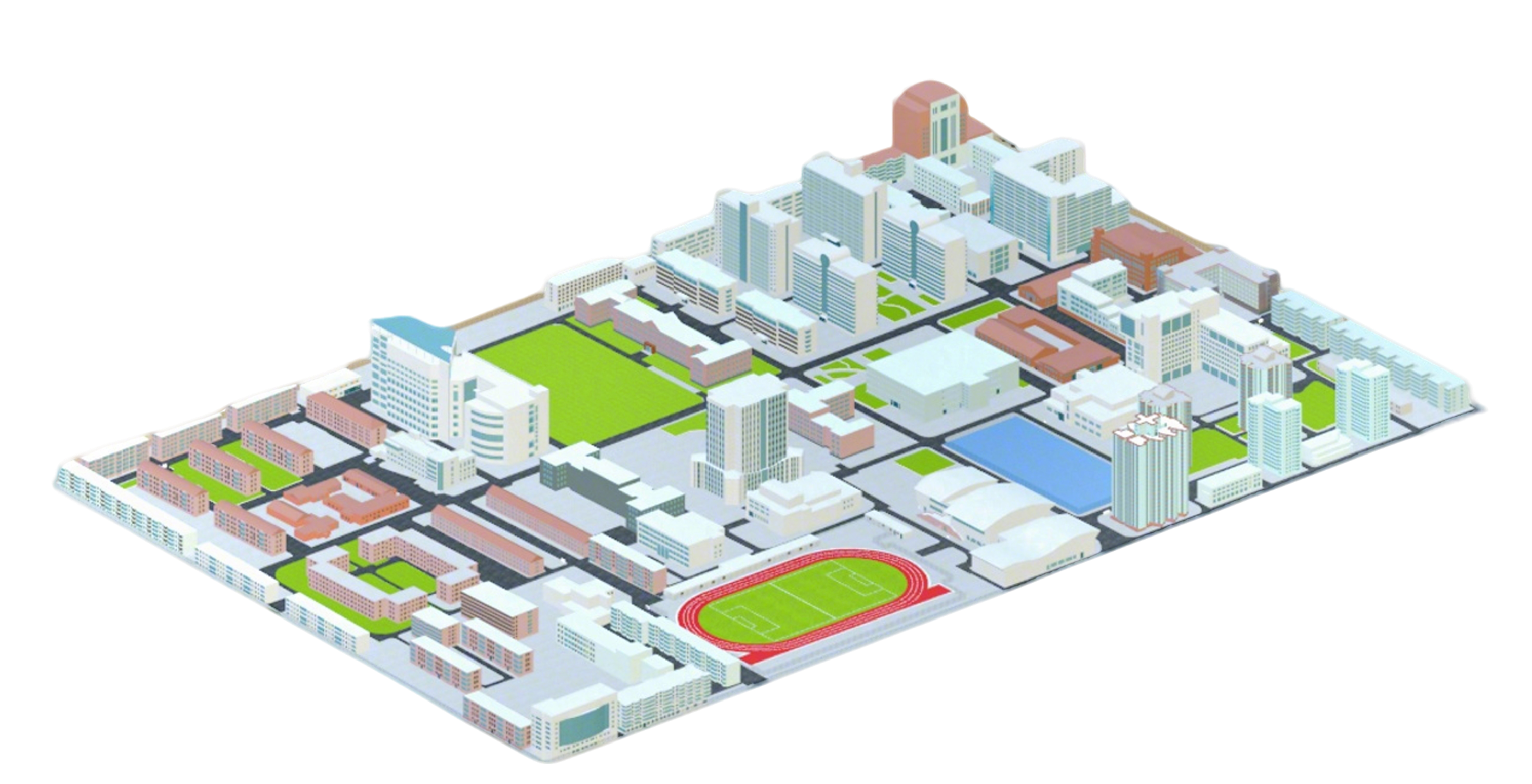}
    \label{fig:fig5a}
}

\vspace{0.25cm}

\subfloat[]{%
    \includegraphics[height=0.16\textheight]{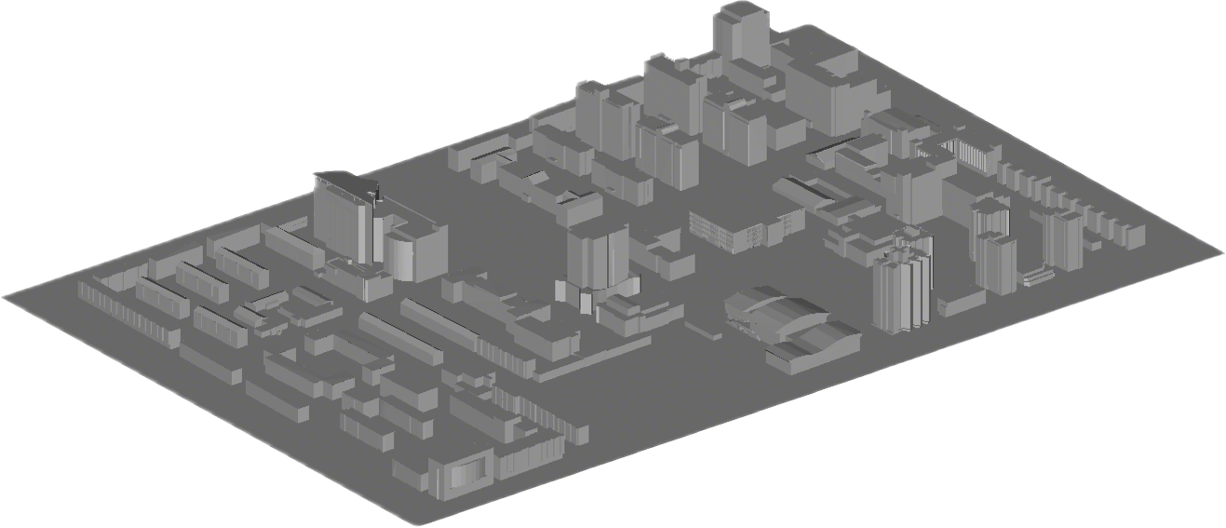}
    \label{fig:fig5b}
}

\caption{Multi-modal dataset construction based on the BUPT campus digital twin. (a) Digital twin scenario. (b) Ray-tracing scenario.}
\label{fig:fig5}
\end{figure}

For the depth prediction task, we adopt the inherited depth loss from the HunyuanWorld-Mirror model. Specifically, the depth loss is defined to enforce consistency between the predicted and ground-truth depth maps, together with a gradient-based term to preserve geometric structures. The depth loss can be expressed as
\begin{equation}
\mathcal{L}_{\mathrm{dep}}
=
\mathcal{L}_{\mathrm{reg}}
+
\beta\,\mathcal{L}_{\mathrm{grad}},
\end{equation}
where \(\mathcal{L}_{\mathrm{reg}}\) denotes the valid-pixel depth regression term, \(\mathcal{L}_{\mathrm{grad}}\) denotes the gradient-based consistency term, and \(\beta\) is the weighting coefficient. This formulation encourages accurate depth recovery while preserving local geometric continuity. 

The above objective is beneficial because it enables the proposed framework to jointly optimize CSI prediction and depth prediction within a unified training process. As a result, the proposed optimization strategy improves both channel prediction performance and environment reconstruction quality.

\section{SIMULATION SETUP AND \\PERFORMANCE ANALYSIS}\label{sec:simulation}
\subsection{Multi-Modal Dataset Construction}

To implement and evaluate the proposed EMWM for joint 3D environment reconstruction and channel prediction, we construct a large-scale multi-modal dataset based on the digital twin of the BUPT campus, as illustrated in \Cref{fig:fig5a}.
Specifically, following a similar approach to the BUPTCMCC-DataAI-6G \cite{ref34, ref35} dataset, the real campus environment is first scanned by a laser scanner to obtain accurate geometric information of buildings, roads, and surrounding infrastructures. 
Based on the acquired scanning data, a digital twin scene of the BUPT campus is reconstructed and imported into the autonomous driving simulator CARLA \cite{ref36}, which preserves the spatial layout and road topology of the real environment.

In the digital twin scenario, 36 predefined routes are designed to cover the major road segments within the campus. 
A mobile vehicle travels along these routes to collect synchronized sensing data. 
Instead of recording data at fixed time intervals, the data are collected according to the traveled distance of the vehicle along the route. 
Specifically, one sample is recorded whenever the vehicle moves forward by 0.1 m along the route, thereby ensuring uniform spatial sampling density across the entire dataset.

At each sampling position, the sensing system mounted on the vehicle captures multi-view RGB images and corresponding depth data simultaneously. 
Specifically, \(N_c = 6\) cameras are installed on the vehicle to cover six surrounding directions, namely, front, right-front, right-behind, behind, left-behind, and left-front, thereby achieving full \(360^\circ\) perception. 
Each camera captures images with a resolution of \(W \times H = 518 \times 518\) and a field of view of \(90^\circ\). 
The cameras are mounted at a height of 1.8 m, which provides a realistic sensing perspective for driving scenarios. 
The corresponding camera  parameters are determined based on the resolution and field of view, and are later utilized for geometric back-projection.

Furthermore, point clouds are generated from the collected multi-view depth images following the point cloud generation method introduced in Section~\ref{sec:framework}. 
For each sampling position, the point clouds from all six viewpoints are aggregated to form a  3D representation of the surrounding environment.

After completing visual and depth data collection in CARLA, the digital twin scene is further imported into the ray-tracing software Wireless InSite \cite{ref37} for channel simulation, as illustrated in \Cref{fig:fig5b}. 
The BS is deployed on the rooftop of the tallest building, serving as a transmitter to ensure coverage across the campus area.
Then, the receiver positions collected along the vehicle trajectories are injected into the ray-tracing simulator, from which the corresponding multipath propagation information is generated for each sampling point and further used to compute the channel responses.

The detailed ray-tracing parameters are summarized in Table~\ref{tab:raytracing}. 
Specifically, the carrier frequency is set to 6.5 GHz, and the X3D model is adopted to achieve high-accuracy full 3D propagation modeling. 
The BS is equipped with \(N_b = 64\) antennas with half-wavelength spacing, while the OFDM system employs \(N_s = 64\) subcarriers. 
In addition, up to 15 propagation paths are considered for each receiver point, with a maximum reflection order of 6 and diffraction order of 1. 
These configurations ensure that the simulated channels accurately capture the multipath propagation characteristics in complex urban environments.
\begin{table}[!t]
    \centering
    \caption{Ray Tracing Parameters}
    \label{tab:raytracing}
    \newlength{\colwidth}
    \setlength{\colwidth}{\dimexpr0.235\textwidth-2\tabcolsep\relax}
    \begin{tabular}{|>{\centering\arraybackslash}p{\colwidth}|>{\centering\arraybackslash}p{\colwidth}|}
        \hline
        Parameter & Value \\
        \hline
        Carrier Frequency & 6.5 GHz \\
        \hline
        Propagation Model & X3D \\
        \hline
        BS Antennas & 64 \\
        \hline
        Antenna Interval & 0.5 wave-length \\
        \hline
        OFDM Subcarriers & 64 \\
        \hline
        Reflection Order & 6 \\
        \hline
        Diffraction Order & 1 \\
        \hline
        Paths Per Receiver Point & 15 \\
        \hline
    \end{tabular}
\end{table}

Through the above procedures, a total of 36,197 sampling positions are collected over the 36 predefined routes, resulting in a unified multi-modal dataset. 
Each sample contains the vehicle position, multi-view RGB images, depth maps, reference point clouds, and the corresponding complete CSI, all associated with the same sampling location. 
The resulting dataset provides a consistent basis for training and evaluating the proposed EMWM architecture on joint environment reconstruction and channel prediction.

\subsection{Experiment Setup}
\subsubsection{Network and Training Parameters}
The constructed dataset is divided into training, validation, and test sets with a ratio of 7:1:2, where the data collected from the 36 predefined routes are randomly shuffled before partitioning. For the CSI input, the numbers of sampled antennas and subcarriers are set to \(N_b^{p} = 32\) and \(N_s^{p} = 32\), respectively, corresponding to using 25\% of the total pilot resources. For the visual input, each input image is partitioned into \(N_p = 1369\) patches. The embedding dimension of the world-model backbone is set to \(d_m = 1024\). To improve parameter efficiency during fine-tuning, the LoRA strategy is employed. Specifically, the rank parameter is set to \(r = 8\), and the scaling factor is set to \(\alpha = 32\), which are applied to the query, key, and value matrices within the multi-head attention modules. In addition, the MoE-based CSI prediction head adopts \(K = 5\) experts to enhance the representation capability for complex wireless channel prediction. For the joint optimization objective, the weighting coefficients for CSI prediction loss and depth prediction loss are set to \(\lambda_{\mathrm{csi}} = 1\) and \(\lambda_{\mathrm{dep}} = 1\), respectively, ensuring balanced supervision between two tasks. Moreover, the gradient-based consistency term is weighted by a coefficient \(\beta = 0.2\) to preserve geometric consistency in depth prediction. The proposed framework is trained using the Adam optimizer with a learning rate of \(1\times10^{-4}\). The training process is conducted for 1 epoch with a batch size determined by hardware constraints. The entire training process is conducted on two NVIDIA GeForce RTX 4090 GPUs.

\begin{figure}[!t]
\centering
\includegraphics[width=\linewidth]{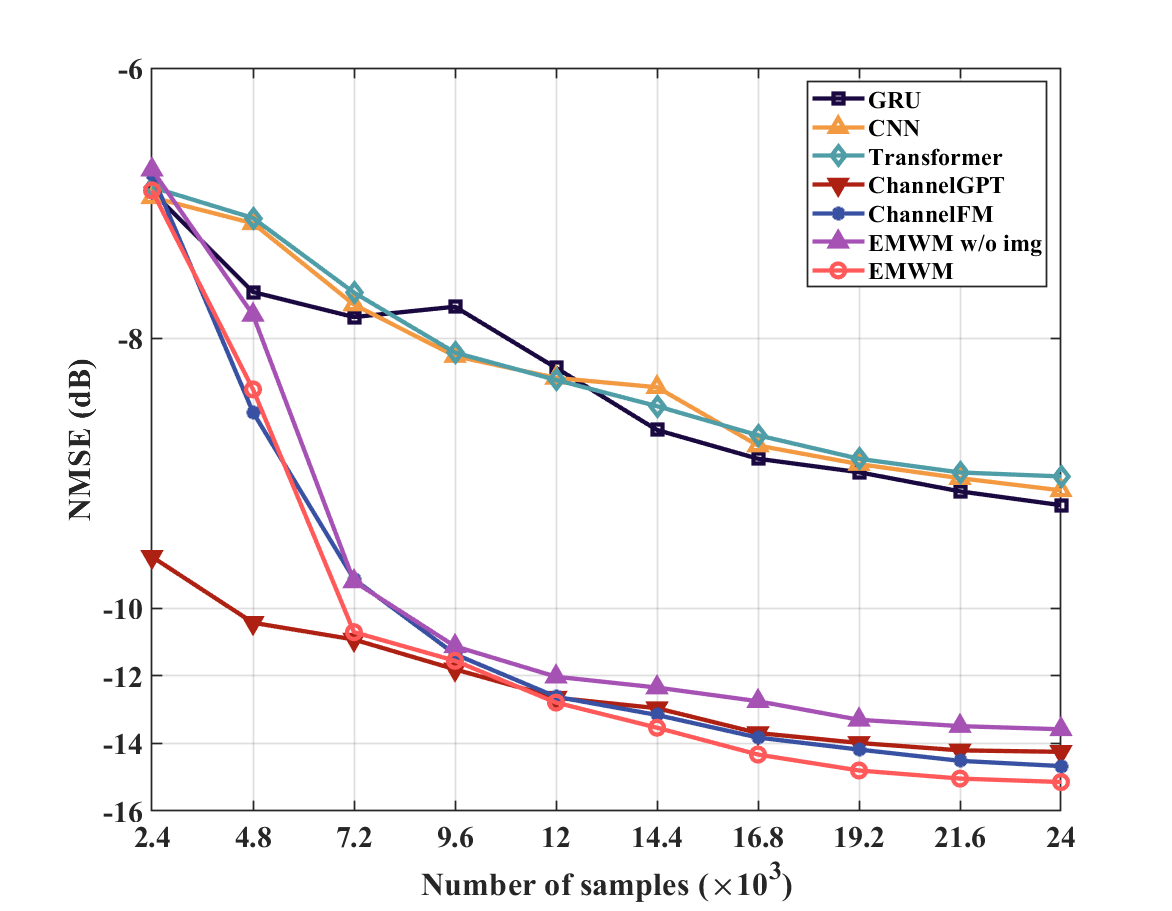}
\caption{Validation NMSE versus the number of training samples.}
\label{fig:nmse_curve}
\end{figure}

\begin{figure}[!t]
\centering
\includegraphics[width=\linewidth]{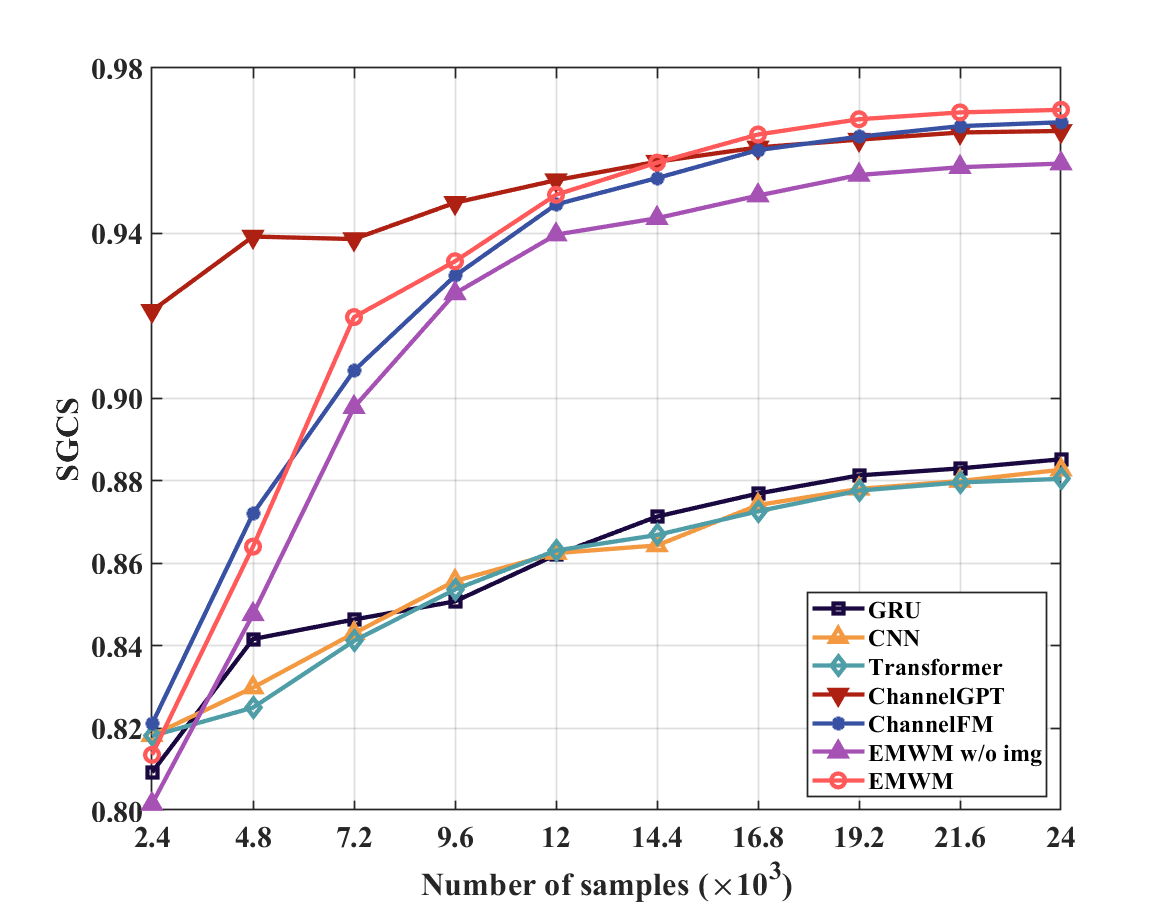}
\caption{Validation SGCS versus the number of training samples.}
\label{fig:sgcs_curve}
\end{figure}
\begin{figure}[!t]
\centering
\includegraphics[width=\linewidth]{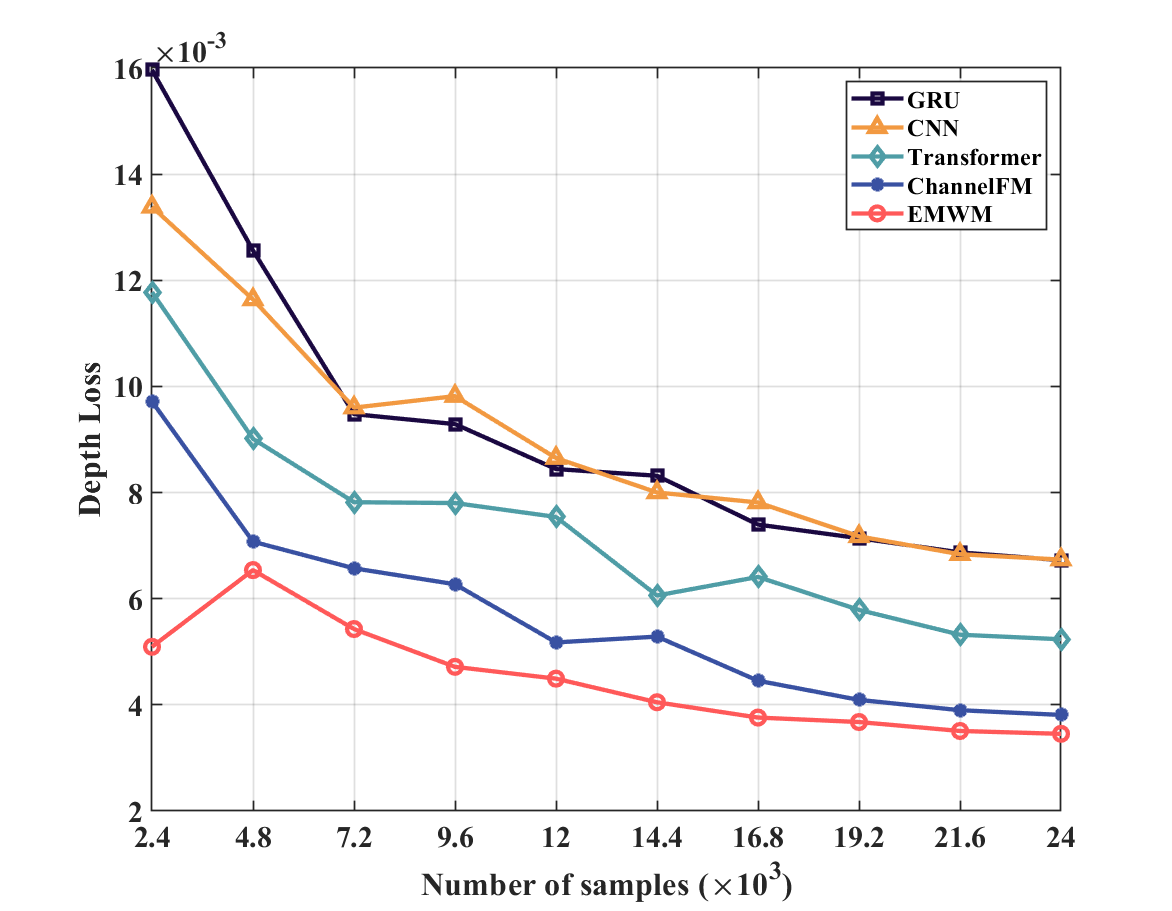}
\caption{Validation depth loss versus the number of training samples.}
\label{fig:depth_loss_curve}
\end{figure}

\begin{figure}[!t]
\centering
\includegraphics[width=\linewidth]{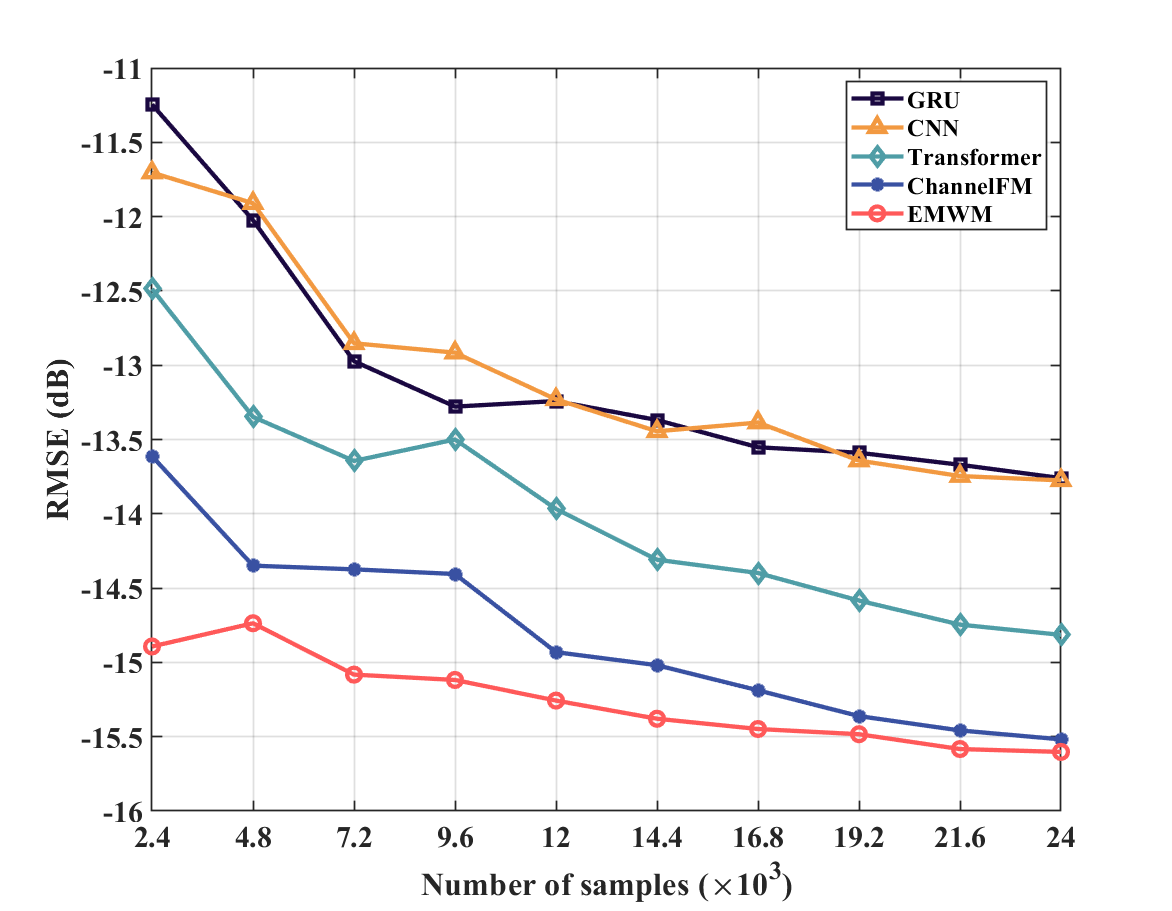}
\caption{Validation depth RMSE versus the number of training samples.}
\label{fig:rmse_curve}
\end{figure}

\subsubsection{Baselines}

To validate the effectiveness of the proposed  EMWM architecture, we implement five representative baselines, including three conventional deep learning architectures and two LLM-based methods. 
Among them, GRU, CNN, Transformer, and ChannelFM are implemented as joint-task baselines by replacing the world-model backbone of EMWM. 
For a fair comparison, these four baselines employ the same visual and CSI tokenization modules, depth prediction head, MoE-based CSI prediction head, joint loss functions, and training settings as EMWM. 
In addition, ChannelGPT is introduced as a multi-modal single-task baseline that takes multi-view images and partial CSI as inputs but predicts only the complete CSI. 

\textbullet \textbf{GRU:} 
The GRU baseline adopts a gated recurrent unit to model dependencies across the token sequence. 
The tokens are first normalized and then processed by a single-layer GRU with hidden size 2048 and dropout 0.1. 
The resulting features are subsequently fed into the CSI prediction head and depth prediction head for downstream tasks.

\textbullet \textbf{CNN:} 
The CNN baseline replaces the backbone with a convolutional mapping applied along the token dimension. 
Specifically, the token sequence is reshaped and processed by a 1D convolutional network composed of a $3\times1$ convolution followed by GELU activation and a $1\times1$ convolution.

\textbullet \textbf{Transformer:} 
The Transformer baseline employs a lightweight Transformer encoder to replace the world-model backbone. 
The token sequence is first linearly projected and then processed by a single-layer Transformer encoder with multi-head self-attention, where the embedding dimension is set to 2048 with 8 attention heads.

\textbullet \textbf{ChannelFM \cite{ref39}:} 
The ChannelFM  baseline adopts the 1B version of Llama3.2 \cite{ref38} as the backbone. 
The visual tokens and CSI tokens are projected into the hidden space with embedding dimension 2048 and then processed by the transformer blocks of Llama.

\textbullet \textbf{ChannelGPT \cite{ref17}:}
The ChannelGPT baseline takes multi-view images and partial CSI as its inputs. The images and partial CSI are transformed into visual tokens and CSI tokens, respectively, and jointly processed by a GPT-2 backbone. 
The resulting fused token representations are then combined with the  partial CSI through a proximal-style reconstruction head to predict the complete CSI.

\begin{figure*}[!t]
\centering
\includegraphics[width=0.98\textwidth]{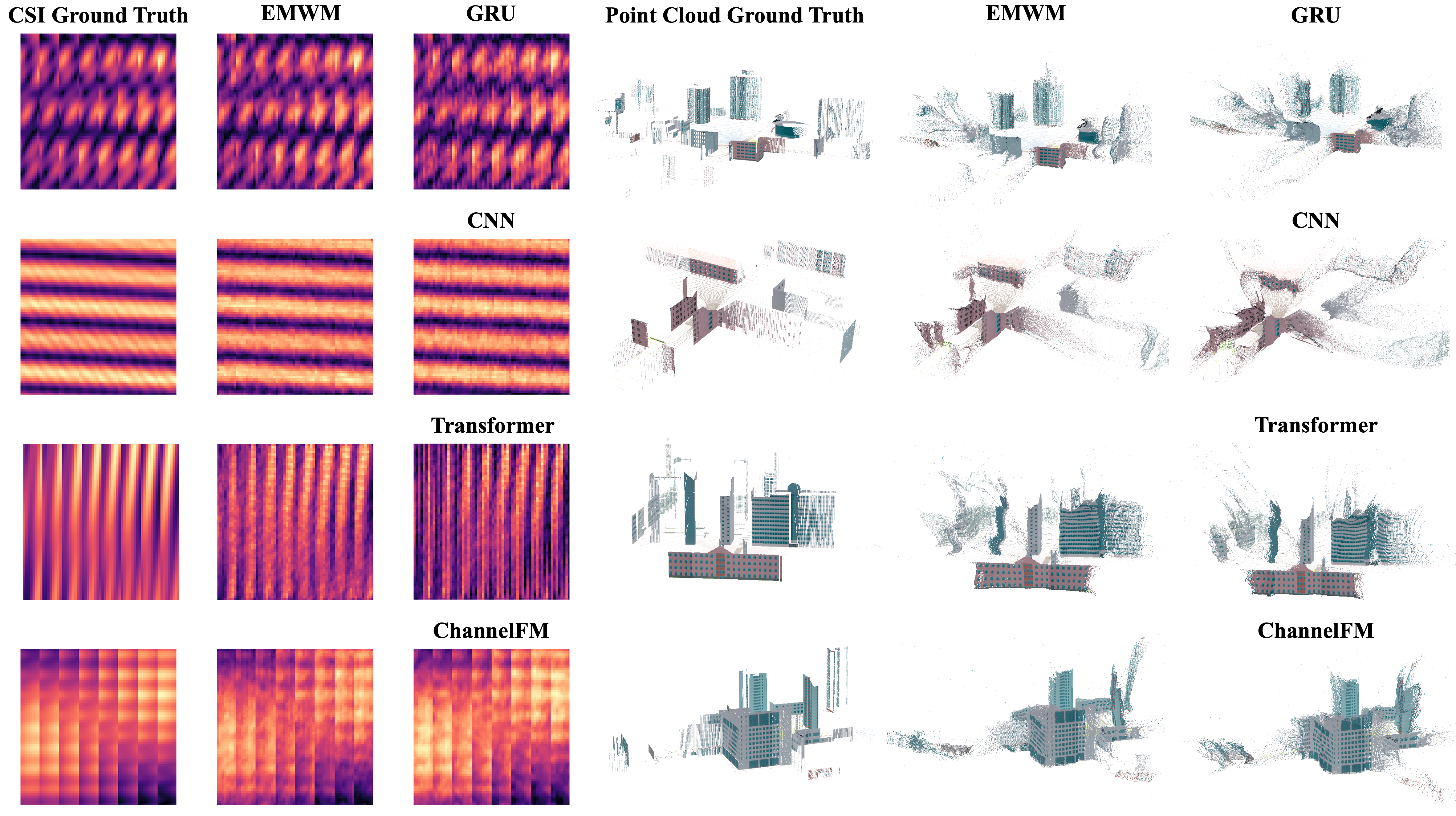}
\caption{Qualitative comparison between the predicted results and ground truth for CSI prediction and environment reconstruction.}
\label{fig:visualization}
\end{figure*}

\subsection{Performance Analysis}
\subsubsection{Prediction Accuracy}We first evaluate the prediction performance from the perspectives of CSI prediction and environment reconstruction. For CSI prediction, NMSE is adopted to measure the normalized reconstruction error, while the squared generalized cosine similarity (SGCS) is further employed to evaluate the structural consistency between the predicted CSI and the ground-truth CSI. Let $\mathbf{H}_{m,n}$ and $\widehat{\mathbf{H}}_{m,n}$ denote the ground-truth and predicted CSI coefficients at the $m$-th antenna and the $n$-th subcarrier, respectively. The NMSE is defined as
\begin{equation}
\mathrm{NMSE}
=
\frac{
\sum_{m=1}^{N_b}\sum_{n=1}^{N_s}
\left|
\mathbf{H}_{m,n}
-
\widehat{\mathbf{H}}_{m,n}
\right|^2
}{
\sum_{m=1}^{N_b}\sum_{n=1}^{N_s}
\left|
\mathbf{H}_{m,n}
\right|^2
}.
\end{equation}

The SGCS is defined as
\begin{equation}
\mathrm{SGCS}
=
\frac{
\left|
\sum_{m=1}^{N_b}
\sum_{n=1}^{N_s}
\widehat{\mathbf{H}}_{m,n}^{*}
\mathbf{H}_{m,n}
\right|^2
}{
\left(
\sum_{m=1}^{N_b}
\sum_{n=1}^{N_s}
\left|
\widehat{\mathbf{H}}_{m,n}
\right|^2
\right)
\left(
\sum_{m=1}^{N_b}
\sum_{n=1}^{N_s}
\left|
\mathbf{H}_{m,n}
\right|^2
\right)
},
\end{equation}
where $(\cdot)^*$ denotes the conjugate operation.

For environment reconstruction, root mean squared error (RMSE) is employed to quantify the pixel-wise discrepancy between the predicted and ground-truth depth maps. Let $\mathbf{D}_i(w,h)$ and $\widehat{\mathbf{D}}_i(w,h)$ denote the ground-truth and predicted depth values at pixel location $(w,h)$ of the $i$-th camera viewpoint, respectively. The RMSE is defined as
\begin{equation}
\mathrm{RMSE}
=
\sqrt{
\frac{1}{N_cWH}
\sum_{i=1}^{N_c}
\sum_{w=1}^{W}
\sum_{h=1}^{H}
\left(
\widehat{\mathbf{D}}_{i}(w,h)
-
\mathbf{D}_{i}(w,h)
\right)^2
}.
\end{equation}

As shown in Fig.~\ref{fig:nmse_curve} and Fig.~\ref{fig:sgcs_curve}, we compare the validation NMSE and SGCS of different methods during training as the number of training samples increases.  It can be observed that EMWM consistently achieves lower NMSE than GRU, CNN, and Transformer, demonstrating that the world-model-based backbone provides stronger representation capability than conventional neural networks. Compared with  ChannelGPT and ChannelFM, EMWM also obtains a lower final NMSE, which indicates that the world model with visual-geometric priors is more suitable than an LLM backbone mainly designed for sequential text modeling. In addition, EMWM consistently outperforms EMWM w/o img throughout training, showing that multi-view images provide complementary environmental information for improving CSI prediction. In terms of SGCS, EMWM achieves the highest final similarity and reaches approximately 0.97. This confirms that the proposed method not only reduces the CSI reconstruction error, but also better preserves the overall structure of the channel.

For environment reconstruction, we compare the validation depth loss and RMSE of the joint-task methods during training, as shown in Fig.~\ref{fig:depth_loss_curve} and Fig.~\ref{fig:rmse_curve}. EMWM consistently achieves the lowest depth loss over the whole training process and converges to approximately 0.0035.  Meanwhile, it also obtains the lowest RMSE  among all methods, demonstrating superior depth prediction accuracy. Compared with the conventional backbones, ChannelFM provides a clear performance improvement, but EMWM further reduces the error by exploiting the geometric representation capability inherited from the world model. These results verify that the proposed world-model based framework is effective in improving depth prediction accuracy.

As shown in Fig.~\ref{fig:visualization}, we provide qualitative comparisons of the predicted CSI and reconstructed 3D point clouds for different methods. For CSI prediction, the result generated by EMWM is visually closest to the ground truth, whereas the baseline results exhibit more noticeable distortions. For environment reconstruction, the point clouds generated by EMWM retain clearer building contours and more complete facade structures than those produced by the compared baselines. In contrast, the baseline point clouds contain more pronounced geometric deformation and structural incompleteness. These qualitative results are consistent with the quantitative evaluation and demonstrate the effectiveness of EMWM in jointly performing 3D environment reconstruction and CSI prediction.

Table~\ref{tab:test_results} summarizes the final test performance of different methods. It can be observed that EMWM achieves the best results across all evaluation metrics. This observation is highly consistent with the trends shown by the validation curves. Therefore, the agreement between the validation results and the test results verifies the effectiveness of the training process and demonstrates that the proposed framework maintains stable prediction performance on unseen test samples. Together with the qualitative visualization, these results confirm the prediction accuracy of EMWM for both environment reconstruction and channel prediction.
\begin{table}[htbp]
\centering
\caption{Test Results of Different Methods}
\label{tab:test_results}
\renewcommand{\arraystretch}{1.12}
\setlength{\tabcolsep}{0pt}
\begin{tabular*}{\columnwidth}{@{\extracolsep{\fill}}lcccc}
\hline
\\[-2.5ex]
Method & NMSE (dB) & SGCS & Loss ($\times 10^{-3}$ ) & RMSE (dB)\\
\hline
GRU & -9.19 & 0.8847 & 6.6816 & -13.75 \\
CNN & -9.08 & 0.8820 & 6.7064 & -13.77 \\
Transformer & -9.00 & 0.8803 & 5.2287 & -14.83 \\
ChannelFM & -14.84 & 0.9679 & 3.7572 & -15.56 \\
EMWM & \textbf{-15.15} & \textbf{0.9699} & \textbf{3.4158} & \textbf{-15.64} \\
\hline
\end{tabular*}
\end{table}
\begin{figure}[!b]
\centering
\includegraphics[width=\linewidth]{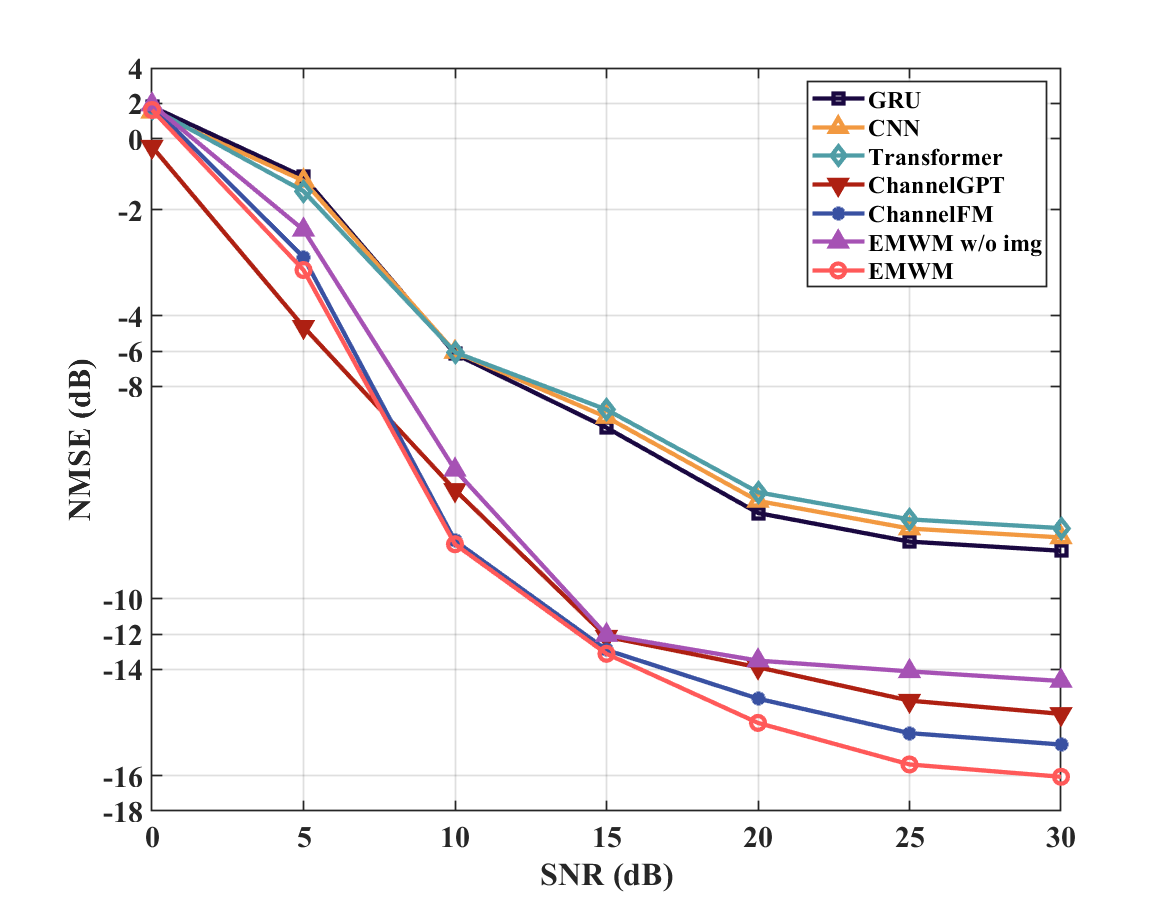}
\caption{NMSE performance under different SNR conditions.}
\label{fig:snr_nmse}
\end{figure}
\begin{figure}[!b]
\centering
\includegraphics[width=\linewidth]{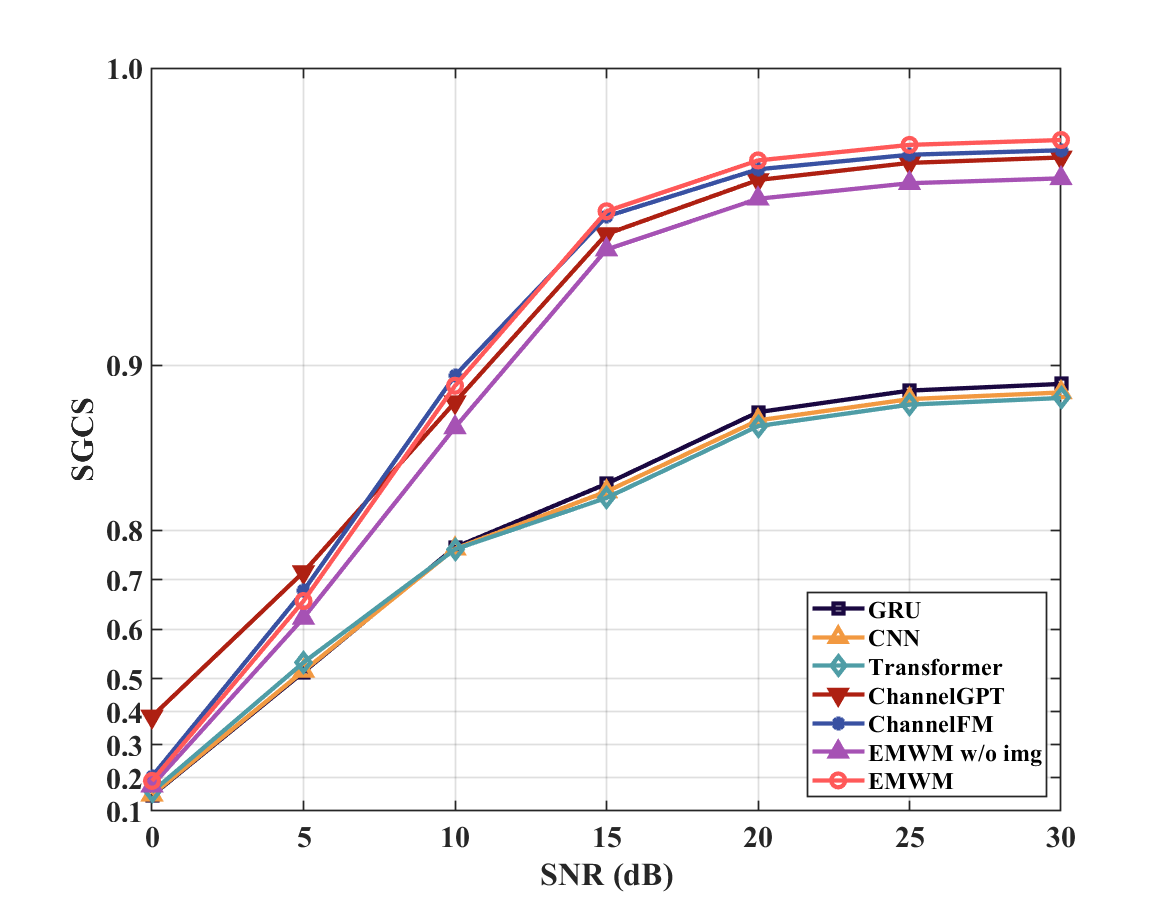}
\caption{SGCS performance under different SNR conditions.}
\label{fig:snr_sgcs}
\end{figure}
\subsubsection{Robustness Against Noise}

In practical wireless systems, the acquired partial CSI is inevitably affected by channel estimation errors and receiver noise. Therefore, the robustness against CSI noise is important for evaluating whether a prediction model can maintain reliable performance under imperfect input conditions. In the above experiments, all methods are trained under the SNR condition of 20 dB.  To further verify the robustness of different methods, we perform an additional robustness evaluation on the validation set under different SNR conditions ranging from 0 dB to 30 dB, and compare the corresponding NMSE and SGCS results.

As shown in Fig.~\ref{fig:snr_nmse}, the NMSE of all methods decreases as the SNR increases, which is reasonable since a higher SNR provides more accurate partial CSI inputs. When the SNR is low, the prediction performance of all methods is degraded due to the strong noise corruption in the CSI input.  As the SNR increases, EMWM shows a clearer advantage over GRU, CNN, and Transformer and achieves  lower NMSE in most testing conditions. Compared with ChannelGPT and ChannelFM, EMWM also obtains better NMSE performance at medium and high SNRs. Moreover, EMWM consistently outperforms EMWM w/o img across all considered SNR conditions, indicating that multi-view images contribute to improving the robustness of CSI prediction.

Fig.~\ref{fig:snr_sgcs} further presents the SGCS performance under different SNR conditions. It can be observed that the SGCS of all methods increases with the SNR, indicating that cleaner CSI  helps preserve the structural information of the channel. Notably, when the SNR is 15 dB, EMWM already achieves a higher SGCS than GRU, CNN, and Transformer at 30 dB. This indicates that the proposed method can maintain strong structural consistency even under relatively noisy CSI inputs. These results further demonstrate that EMWM maintains robust CSI prediction performance under imperfect input conditions.

\begin{figure}[!t]
\centering
\includegraphics[width=\linewidth]{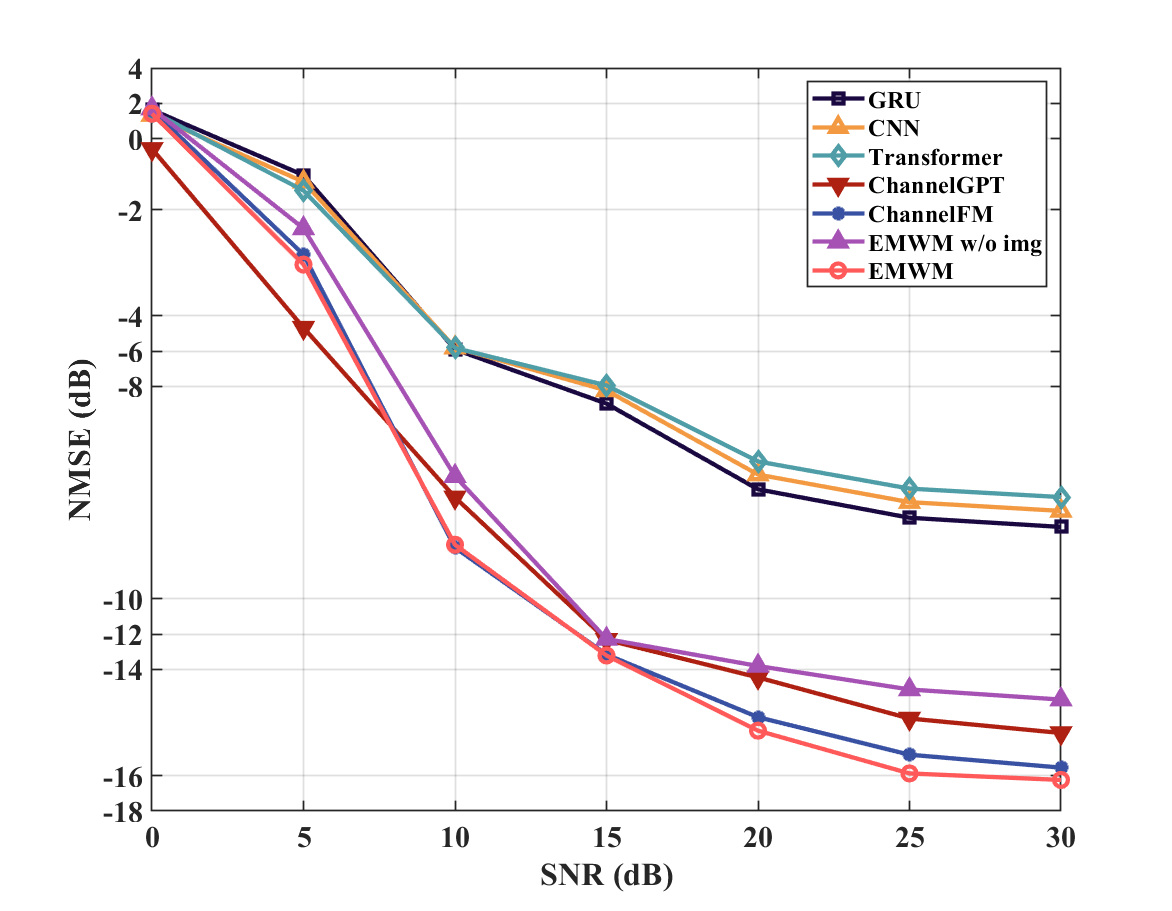}
\caption{Cross-frequency zero-shot NMSE performance at 28 GHz under different SNR conditions.}
\label{fig:cross_frequency_nmse}
\end{figure}

\subsubsection{Cross-Frequency Zero-Shot Generalization}

Generalization across carrier frequencies is important for evaluating whether a prediction model can adapt to new propagation conditions without retraining. To examine the zero-shot generalization capability, we change the ray-tracing carrier frequency from 6.5 GHz to 28 GHz and directly use the 28 GHz validation data under different SNR conditions as the model input, while keeping the trained models unchanged.

As shown in Fig.~\ref{fig:cross_frequency_nmse}, the NMSE of different methods is compared under SNRs ranging from 0 dB to 30 dB. It can be observed that the conventional models suffer from performance degradation at 28 GHz compared with the 6.5 GHz setting. For example, at 20 dB, the NMSE values of GRU, CNN, and Transformer degrade from -9.19 dB, -9.08 dB, and -9.00 dB to -8.97 dB, -8.83 dB, and -8.71 dB, respectively. In contrast, ChannelGPT, ChannelFM, EMWM w/o img, and EMWM achieve slightly lower NMSE at 28 GHz. Specifically, the NMSE of ChannelGPT improves from -13.87 dB to -14.15 dB, while that of ChannelFM improves from -14.55 dB to -14.90 dB. Similarly, the NMSE values of EMWM w/o img  and EMWM improve from -13.50 dB and -15.01 dB to -13.80 dB and -15.15 dB, respectively. These results indicate that the large-model-based methods maintain strong  performance under the unseen 28 GHz carrier frequency, demonstrating their cross-frequency zero-shot generalization capability.

\section{Conclusion}\label{sec:conclusion}
In this paper, we propose EMWM for 6G as a unified world-model paradigm that jointly models the physical environment and  wireless channel. Under this paradigm, we develop a concrete EMWM architecture for joint environment reconstruction and channel prediction. By taking partial CSI and multi-view images as heterogeneous electromagnetic observations, the EMWM architecture simultaneously reconstructs complete CSI and predicts multi-view depth maps, which are further converted into 3D point clouds. To effectively integrate radio-frequency and optical information, CSI tokens are incorporated into the world-model backbone together with visual tokens, while an MoE-based CSI prediction head and a depth prediction head are employed for task-specific outputs. Moreover, a large-scale multi-modal dataset is constructed based on the BUPT campus digital twin, including synchronized multi-view images, depth maps, point clouds, and CSI. Experimental results demonstrate that EMWM outperforms conventional neural networks and LLM-based methods in both CSI prediction and 3D environment reconstruction. In addition, evaluations under different SNR conditions and cross-frequency zero-shot validation at 28 GHz further verify its robustness under imperfect CSI inputs and its generalization capability across carrier frequencies. These results validate the effectiveness of EMWM and indicate its potential as a common modeling foundation for diverse 6G tasks and application scenarios.

\bibliographystyle{IEEEtran}
\bibliography{ref}
\end{document}